\documentclass[referee,sn-basic]{sn-jnl}

\usepackage{graphicx}%
\usepackage{multirow}%
\usepackage{amsmath,amssymb,amsfonts}%
\usepackage{amsthm}%
\usepackage{mathrsfs}%
\usepackage[title]{appendix}%
\usepackage{xcolor}%
\usepackage{textcomp}%
\usepackage{manyfoot}%
\usepackage{booktabs}%
\usepackage{algorithm}%
\usepackage{algorithmicx}%
\usepackage{algpseudocode}%
\usepackage{listings}%
\usepackage{todonotes}
\usepackage{array}

\newcommand{\Dumux}{DuMu$^x$}

\usepackage[version=3,arrows=pgf-filled ]{ mhchem}

\usepackage[final]{changes}
\usepackage{xcolor}
\usepackage[normalem]{ulem} 

\renewcommand{\added}[2][]{\textcolor{red}{#2}}
\renewcommand{\deleted}[2][]{\textcolor{red}{\sout{#2}}}
\renewcommand{\replaced}[3][]{\textcolor{red}{#2} \textcolor{gray}{\sout{#3}}}

\renewcommand{\added}[2][]{#2}
\renewcommand{\deleted}[2][]
{}
\renewcommand{\replaced}[3][]{#2}

\theoremstyle{thmstyleone}%
\theoremstyle{thmstyletwo}%

\theoremstyle{thmstylethree}%

\begin{document}

\title[Article Title]{A hydro-geomechanical porous-media model to study effects of engineered carbonate precipitation in faults}


\author*[1]{\fnm{Yue} \sur{Wang}}\email{yue.wang@iws.uni-stuttgart.de}

\author[1]{\fnm{Holger} \sur{Class}}\email{holger.class@iws.uni-stuttgart.de}
\equalcont{These authors contributed equally to this work.}

\affil*[1]{\orgdiv{Department of Hydromechanics and Modelling of Hydrosystems}, \orgname{ Institute for Modelling Hydraulic and Environmental Systems, University of Stuttgart}, \orgaddress{ \city{Stuttgart}, \postcode{70569}, \country{Germany}}}

    \abstract{Hydro-geomechanical models are required to predict or understand the impact of subsurface engineering applications as, for example, in gas storage in geological formations. This study puts a focus on engineered carbonate precipitation through biomineralization in a fault zone of a cap-rock to reduce gas leakage from a reservoir. Besides hydraulic properties like porosity and permeability, precipitated carbonates also change the mechanical properties of the rock. We present a conceptual modeling approach implemented into the open-source simulator \Dumux and, after verification examples,  at hand of a \ce{CO2}-storage scenario, we discuss impacts of biomineralization on the stress distribution in the rock and potentially altered risks of fault reactivations and induced seismic events. 
    
    The generic study shows the tendency towards increased stiffness due to precipitated carbonate, which may cause shear failure events to occur earlier than in an untreated setup, while the magnitude of the seismicity is smaller. }

\keywords{hydro-geomechanical coupling, biomineralization, fault reactivation, model verification}



\maketitle

\section{Introduction}\label{sec1}

The injection of large amounts of fluids into the subsurface is a necessary and intentional engineering measure in various technologies that are linked to energy production or storage, like in geological carbon sequestration, hydraulic fracturing, geothermal systems, etc. Fluid injection is in some instances also a concomitant feature, for example, when waste-waters from conventional or unconventional hydrocarbon production are re-injected into geological formations. In particular in the liberal societies, these technologies are often controversially discussed in view of various risks, which are related to hydro-geomechanical processes and include, for example, leakage of fluids or induced seismic events \cite{ScheerFlemischClass2021}. 

Leakage from geological reservoirs reduces the efficiency of storage and may pose a threat to the environment. Engineered carbonate precipitation\added{, i.e., the intentional initiation of calcium carbonate  formation (\ce{CaCO3}) through controlled biochemical, chemical, or physical mechanisms,} is discussed as a technology that can be implemented in different ways to seal leakage pathways (  \cite{Kirkland2020, Phillips2018, Phillips2016, Cunningham2014, Cuthbert2013}).
The technology has already successfully made its way from the small-scale test phase to field-scale application \cite{Cunningham2019}. From other applications related to ground improvement, it is known that induced carbonate precipitation can improve also the geomechanical properties of porous media \cite{Zeng2021,Paassen2010}, which is so far rarely considered in the context of the aforementioned intentional sealing of potential leakage pathways.

Engineered carbonate precipitation can be achieved by different methods, most notably the microbially induced carbonate precipitation (MICP) is distinguished from the enzymatically induced carbonate precipitation (EICP). While MICP typically relies on the bacterium Sporosarcina Pasteurii as a source for the enzyme urease, this enzyme is provided in the case of EICP from other sources, for example, from a powder of Jack-Bean meal. Most importantly, urease catalyzes the hydrolysis reaction of urea ($\ce{(NH_2)_2CO}$) into ammonia ($\ce{NH_{3}}$) and carbon dioxide ($\ce{CO_2}$). This reaction increases pH and aqueous solutions of ammonia become alkaline. More alkalinity in the solution shifts the equilibrium of inorganic carbon species towards higher concentrations of carbonate ($\ce{CO_{3}^{2-}}$). Thus, in the presence of calcium ($\ce{Ca^{2+}}$), the precipitation of calcium carbonate ($\ce{CaCO_{3}}$) is promoted.
The overall reaction is:
\begin{equation}
\ce{(NH_2)_2CO} + 2\ce{H_2O} + \ce{Ca^{2+}} \longrightarrow
2\ce{NH_{4}^{+}} + \ce{CaCO_{3}}\downarrow
\label{EICP_chem_eq}
\end{equation}
The precipitated $\ce{CaCO_{3}}$ fills voids in the pore space and alters the mechanical properties of the porous matrix.

\added{
While the primary focus of this study is on induced carbonate precipitation (ICP) for targeted sealing of leakage pathways like open fractures, it is worth noting that natural CO$_2$ mineralization in fractured rock formations has been widely studied and can occur under appropriate conditions. Pilot projects and recent studies, such as those in \cite{nisbet2024carbon}, indicate that mineralization in fractured rock can be very significant, dependent on the rock type, though the precise role of fractures in promoting this process remains complex. As reviewed in \cite{kim2023review}, factors such as porosity and permeabilityplay crucial roles in CO$_2$ mineralization and alteration of mechanical strength is noted though not transferred into models so far. } 

Fluid injections into deep subsurface formations affect both the fluid pressures and the stresses and deformations in geological layers. It is important to understand the hydro-geomechanical couplings in order to assess and predict potential impacts and risks. 
One of the scenarios that need to be considered in such applications is concerned with the reactivation of existing faults through shear failure; the literature provides here a number of generic studies which refer to a realistic field-scale application (\cite{Beck-stressdrop,lei2017fault,vil2015geologic,rut2013modeling,cap2011modeling,sib1985note}). In some cases, fault reactivation through fluid injection can lead to measureable, even to serious seismic events (\cite{ell2013injection,fro2014may,hol2013earthquakes,maz2012induced,sha2009fluid,ker2013potentially}).

Numerical simulation of coupled processes is complex and implementations of hydro-geomechanical models are often discussed with respect to numerical accuracy and computational efficiency, since they are typically realized in some kind of appropriate coupling scheme (\cite{Beck2020,kim2012numerical,kim2010sequential,rut2008coupled,rin2014geomechanical}).
In this study, we present an implementation of a coupled hydro-geomechanical model in the numerical simulator \Dumux (\url{www.dumux.org}) \cite{Kochetal2020Dumux,dumux-handbook,Beck-stressdrop,Beck2020}. The model is based on the approach introduced recently by \cite{Beck-stressdrop,Beck2020} in an older version of \Dumux; it is now newly adapted to the current architecture of \Dumux and also further developed regarding the conceptual model. 
In contrast to the model developed by Loyola et al.\cite{Loyola2024}, which extends the non-dimensional fracture model in \Dumux, that accounts for unevenly developed single-phase flow in fractured areas and uses an upscaling method, our approach maintains an equidimensional description of the fault zone. This allows for the incorporation of mechanical property variations, as the mechanical effects in sub-dimensional regions are ignored. This decision is based on practical considerations for addressing large-scale problems.
We introduce below the basic model concept including the governing equations, and the numerical coupling methods. We extended the approach for the first time to include effects of carbonate precipitation, and we discuss how rock-failure criteria are taken into account.
In addition to the improved conceptual model, we amended further the assembly process by integrating caching techniques into the multi-domain assembler, thereby accelerating its performance and enabling the implementation of the incremental stress formulation.

The model is verified at hand of problems from the related literature regarding the aspects of coupling flow and geomechanics as well as the aspects on biomineralization and its implication for geomechanical parameters. \added{In contrast to most hydro-geomechanical simulators, our approach explicitly incorporates effects of precipitation processes in the rock matrix and their mechanical effects. By coupling precipitation-induced changes in porosity and permeability with the material’s mechanical response, the model captures alterations in rock stiffness and strength under evolving geochemical conditions. This goes beyond recent (T)HMC models \cite{wang2024coupled, bhukya2024coupled}, which primarily focus on changes in porosity and rely on empirical relationships to update elastic moduli. In our work, a “cementation” concept is used to mechanistically link precipitation to the rock’s stiffness and strength at the pore scale, thus providing a more robust and scalable basis for reservoir-scale and fault-reactivation scenarios.} Eventually, the main focus of this study is on a reservoir-scale showcase application, where we adapt and extent a fault-reactivation scenario as introduced first by \cite{rut2013modeling} and later adopted and modified by \cite{Beck-stressdrop,Beck2020}. In this study, the showcase scenario is used to investigate how the biomineralization of leakage paths can affect the hydrogeomechanical response of a reservoir. Specifically, we want to see how the sealing acts on the stress state in the reservoir and if continued fluid injection leads to a different pattern of possible failure events. We will discuss this in the context of the uncertainties associated, for example, with initial stress distributions, boundary conditions as well as with model and parameter inaccuracies.

\section{Model concepts}

\subsection{Governing equations}
The flow of multiple fluid phases through a deformable porous medium can be mathematically represented by the mass balance equations of the fluids including Darcy's law as well as the momentum balance of the solid matrix. The equations are coupled via the porosity, i.e., the void space available for fluid, and the pore pressure, i.e., the force exerted on the porous structure by the fluid.

\subsubsection{Mass balance equation of the fluid}
\begin{equation}
    \frac{\partial (\phi_l \varrho_\alpha S_\alpha)}{\partial t} - \text{div} \left\{ \varrho_\alpha \boldsymbol{v_\alpha} \right\} - q_\alpha = 0 , \alpha \in {w,n}.
    \label{eq: 2p-mass-balance}
\end{equation}

$\phi_l = (1+\epsilon_v) \phi$ stands for the absolute void space alteration, while the porosity $\phi$ represents only the relative ratio in the current reference configuration. $\varrho_\alpha$, $S_\alpha$ are respectively the density and the saturation of phase $\alpha$. $\epsilon_v$ is the volume strain. Following the assumption of infinitesimal deformation, $\epsilon_v \approx 0$, so $\phi_l \approx \phi$ in this work.

$v_\alpha$ is the Darcy velocity of phase $\alpha$:
\begin{equation}
    v_\alpha = \frac{k_{r\alpha}}{\mu_\alpha} \mathbf{K} \left(\textbf{grad}\, p_\alpha - \varrho_{\alpha} \mathbf{g} \right),
    \label{eq: Darcys Law}
\end{equation}

and $q_\alpha$ stands for source or sink terms.

Inserting Eq.~\eqref{eq: Darcys Law} into Eq.~\eqref{eq: 2p-mass-balance} returns the mass balance equations for the two phase flow:
\begin{equation}
    \frac{\partial (\phi \varrho_\alpha S_\alpha)}{\partial t} - \text{div} \left\{ \varrho_\alpha \frac{k_{r\alpha}}{\mu_\alpha} \mathbf{K} \left(\textbf{grad}\, p_\alpha - \varrho_{\alpha} \mathbf{g} \right) \right\} - q_\alpha = 0 \;
    \label{eq: 2p-mass-balance-full}
\end{equation}

The nonlinear storage term in Eq.~\eqref{eq: 2p-mass-balance-full} is equivalent to that in the diffusion equation in \cite{cheng:poroelasticity} for single phase flow, where the storage term is linearized with Eq.\eqref{eq: linear porosity law short} and expressed as 
\begin{equation}
    \frac{\partial (
    \phi \varrho_)}{\partial t} = \rho \frac{\partial (\alpha \epsilon + \frac{1}{M} p)}{\partial t},
\label{eq: linear storage term}
\end{equation}
with $\alpha [-]$ as Biot coefficient, $\epsilon [-]$ as volume strain of matrix and $M [Pa]$ as Biot Modulus.

In the subsequent sections, the verification cases, originally employed linear formulations are evaluated against their nonlinear counterparts as implemented within the \Dumux framework.

\subsubsection{Momentum balance of matrix}
Adhering to the commonly assumed quasi-static behavior in the field of modelling geomechanics, the momentum balance can be formulated as:
\begin{equation}
    \nabla\cdot\boldsymbol{\sigma_{\mathrm{eff}}} + \rho \mathbf{g}  = \rho\ddot{\mathbf{u}} \approx 0,
    \label{eq:el-momentum-balance}
\end{equation}
where $\varrho = \sum_\alpha \phi S_\alpha \varrho_\alpha + (1- \phi) \varrho_s$ represents the density of the porous medium and $\boldsymbol{\sigma_\mathrm{eff}}$ is Biot's effective stress, where positive stress indicates tension:

\begin{equation}
    \boldsymbol{\sigma_{\mathrm{eff}}} = \boldsymbol{\sigma} - \alpha p_{\mathrm{eff}} \mathbf{I},
    \label{eq: effective-stress-formulation}
\end{equation}
with $\alpha$ as Biot's effective stress coefficient, $p_{\mathrm{eff}} = \sum S_\alpha p_\alpha$ as effective pore pressure and $\boldsymbol{\sigma}$ the stress tensor after Hookes Law:

\begin{align}
    \boldsymbol{\sigma} &= \lambda\mathrm{tr}(\boldsymbol{\varepsilon}) \mathbf{I} + 2G \boldsymbol{\varepsilon},  \\
    \boldsymbol{\varepsilon} &= \frac{1}{2} \left[ \nabla\mathbf{u} + (\nabla\mathbf{u})^{\mathrm{T}} \right].
    \label{eq: Hookes-Law}
\end{align}

Indeed, the stress tensor $\sigma$ can also be formulated incrementally. In an incremental approach, the changes in stress are calculated step by step (from $\boldsymbol{\sigma}_{prev}$ to $\boldsymbol{\sigma}$), considering the changes in strain ($\Delta u$) and other relevant factors (e.g. variations in bulk modulus). 
\begin{align}
    \boldsymbol{\sigma} &= \boldsymbol{\sigma}_{prev} + \lambda\mathrm{tr}(\boldsymbol{\Delta \varepsilon}) \mathbf{I} + 2G \boldsymbol{\Delta \varepsilon} \\
    \boldsymbol{\Delta \varepsilon} &= \frac{1}{2} \left[ \nabla(\mathbf{\Delta u}) + (\nabla(\mathbf{\Delta u}))^{\mathrm{T}} \right].
    \label{eq: incremental_formulation}
\end{align}
This method is particularly useful for analyzing materials undergoing alterations, such as the matrix under the reaction of induced carbonate precipitation, and complex loading conditions, where providing a precise description of the displacement is often challenging and not feasible.

\begin{align}
    \Delta (\nabla\cdot\boldsymbol{\sigma_{\mathrm{eff}}}) + \Delta (\rho \mathbf{g}) &= 0 \\
    \lambda\mathrm{tr}(\boldsymbol{\Delta \varepsilon}) \mathbf{I} + 2G \boldsymbol{\Delta \varepsilon} + \Delta (\rho \mathbf{g}) &= 0 
\end{align}

Analogous to the incremental stress formulation, the balance equation can also undergo assessment through an incremental approach. This method renders the consideration of momentum equilibrium superfluous for the simulation, with the initial stress exclusively influencing the actual in-situ stress conditions.

In case where it is not explicitly mentioned, the incremental stress formulation and incremental momentum local residual are applied by default.

\subsubsection{Effective porosity}
As aforementioned, the porosity changes are attributable to two factors, the external stress on the solid matrix and the internal pore pressure from contained fluid. 

From the view of micromechanics (proposed in \cite{gassmann1951uber} and represented in \cite{cheng:poroelasticity}) and under the assumption of an ideal porous medium (the unjacketed frame bulk modulus $K_s'$ and the unjacketed pore volume bulk modulus $K_s''$  equal to  $K_s$, the bulk module of skeleton formatting solid), the porosity variation as a mechanical response to the altering stress state is defined as:

\begin{equation}
    \Delta \phi = - \frac{\alpha - \phi}{K} \Delta p_T
    \label{eq: porosity change law}
\end{equation}
where $p_T$ stands for the Terzaghi effective compressive stress, 

\begin{equation}
\Delta p_T = - K \epsilon_v  + \Delta[(\alpha -1) p_{eff}], ,
\label{eq: Terzaghi effective compressive stress}
\end{equation}
where $\epsilon_v$ is the volume strain of the matrix, and $K = \lambda + \frac{2}{3}G$ is the bulk modulus.
The role of pore pressure and Biot coefficient becomes evident when considering the effective compressive stress. In many cases, the Biot coefficient is commonly approximated as $\alpha=1$, thus the compressive stress primarily depends on the deformation of the matrix.

Substituting Eq.\eqref{eq: Terzaghi effective compressive stress} into Eq.\eqref{eq: porosity change law}, we get the effective porosity as 

\begin{align}
    \phi &= \frac{K \phi_0 - \alpha \Delta p}{K-  \Delta p}
    \label{eq: effctive-porosity}
\end{align}

For a given medium consisting of incompressible solids ($\alpha = 1$), Eq.\eqref{eq: effctive-porosity} can be reduced to $\phi = \frac{\phi_0 + \mathrm{div} \mathbf{U}}{1+ \mathrm{div} \mathbf{U}}$ as in \cite{Beck2020}.

A further linearization of Eq.(\ref{eq: porosity change law}), employing the characteristic properties of an ideal porous medium, results in the derivation of the following equation:
\begin{equation}
\Delta \phi = \alpha \epsilon + \frac{1}{M} p_{eff},
\label{eq: linear porosity law short}
\end{equation}
which is identical to the one in \cite{Coussy2003}. For details please refer to the appendix.

Eq.(\ref{eq: porosity change law}) and Eq.(\ref{eq: linear porosity law short}) both align with the incremental stress formulation. However, the latter necessitates an explicit calculation of the effective fluid bulk modulus ($K_f$), which becomes considerably intricate in scenarios of two-phase flow involving capillary pressure. Consequently, the first formulation is implemented, wherein $K_f$ is resolved implicitly. 

\subsubsection{Fixed-stress splitting as decoupling method}
Reviewing the equations for mass and momentum balance, \eqref{eq: 2p-mass-balance-full} and \eqref{eq:el-momentum-balance}, the coupled system can be solved in a sequential manner where the mass balance equation firstly predicts the pore pressure under the assumption of a fixed total stress ($\Delta \varepsilon_v = 0$) between two time steps and the momentum equation then corrects the effective porosity based on the actual pore pressure value. A more detailed algorithm is illustrated in \cite{Beck2020}. However, as pointed out by\cite{split-coefficient}, the algorithm's efficiency is contingent upon the selection of the "bulk modulus", which, although it can be fictitious and nonphysical, is typically assumed to be equal to the actual bulk modulus of the matrix. Beck Yet another advantage is that this assumption often aligns with the boundary conditions in geomechanics simulations, where a fixed stress, rather than absolute deformation, is taken into consideration.

\subsection{Effects of engineered carbonate precipitation on porous media}
Precipitated carbonate minerals in a porous medium alter both hydraulic and mechanical properties of a porous medium. Following the common approach, we use the porosity change to mark the precipitation process. In the following, we will elaborate on how this effect can be qualitatively described.

\subsubsection{Mechanical effects}
The occupation of void pore volume by precipitation is expected to result in a stiffer and more compact elastic matrix. However, it is worth noting that there are currently only few available relationships that can describe this phenomenon. Fauriel and Lalouli \cite{MICP-Model-2012} implemented the linear relationship between shear wave velocity and calcite precipitation, which is observed in the experiment \cite{calcite-wave-speed}. The two elastic moduli can be determined under the assumption of a constant Poisson's ratio. A similar linear relationship is used in the MICP model proposed by X.~Wang (2022) \cite{MICPModelWang2022}.

The linear relationship of this kind is straightforward and evident. Nevertheless, its utility remains constrained by the availability of experimental data. Therefore, we suggest a novel approach that incorporates the effect of carbonate precipitation into the natural process, enhancing its adaptability and flexibility.

Cementation of rock is a natural process, by which sediments or grains are bound together by mineral precipitates, creating a solid mass. This crucial geological process plays a significant role in the formation of sedimentary rocks, contributing to their strength and durability over time. Hence, the additional precipitation introduced by the carbonate precipitation process can be conceptually perceived as an enhanced cementation process. Given the assumption that the rock is composed solely of solid grains and cement, and the precipitation has the same chemical structure as the cement so that they cannot be distinguished from one another, we can use the constant cement model concluded in the work of \cite{Avseth2010} to estimate the mechanical properties of the matrix at each porosity state.

\begin{figure}
    \centering
    \includegraphics[width=0.5\columnwidth]{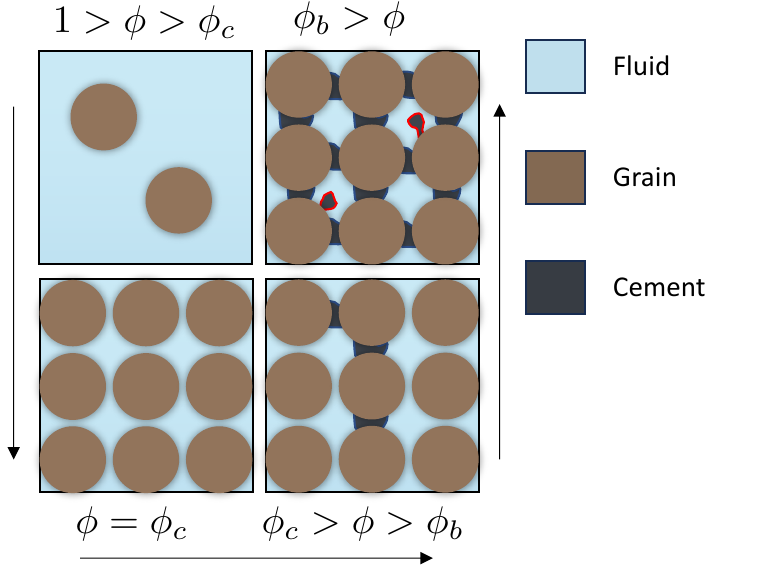}
    \caption{Simplified process (follow subfigures counter clockwise) of rock cementation.}
    \label{fig:Simplifiedprocessofrockcementation}
\end{figure}

The constant cement model represents a hybrid approach, combining insights from both the contact cement model and the friable-sand model proposed by Dvorkin et al \cite{Dvorkin1991, Dvorkin1994}. The contact cement model investigates the influence of cement on the contact points between solid grains. This cement layer acts as an elastic foundation during deformation, which can significantly enhance the elastic modulus. On the other hand, the friable-sand model assesses the pressure-dependent bulk modulus of the matrix at the critical porosity, i.e., the porosity at which solid grains start forming a force-bearing structure, and it utilizes the bounding method to refine the modulus range.  Figure~\ref{fig:Simplifiedprocessofrockcementation} illustrates the rock cementation process, depicting the porosity evolution of the matrix. At extremely high porosity, the sand grains remain suspended in the system, with the fluid being the dominant load-bearing element. As sedimentation progresses and sufficient solid grains accumulate, reaching the critical porosity ($\phi_c$), the skeleton of the matrix stabilizes. Subsequently, the cementation process commences, with precipitation initially occurring at the contact points between grains, leading to a significant increase in the matrix's elastic properties. In the final phase, a well-sorted porosity, denoted as $\phi_b$, emerges, signifying that the contact area has been utilized, and precipitation now occurs randomly throughout the matrix. The mathematical description of the constant cement model is provided in the Appendix. This innovative approach has been subjected to comparison with real rock data in \cite{Avseth2010}, as well as exploration through digital rock simulations in \cite{Wetzel2021_mechanical}.

\begin{figure}
    \centering
    \includegraphics[width=0.7\columnwidth]{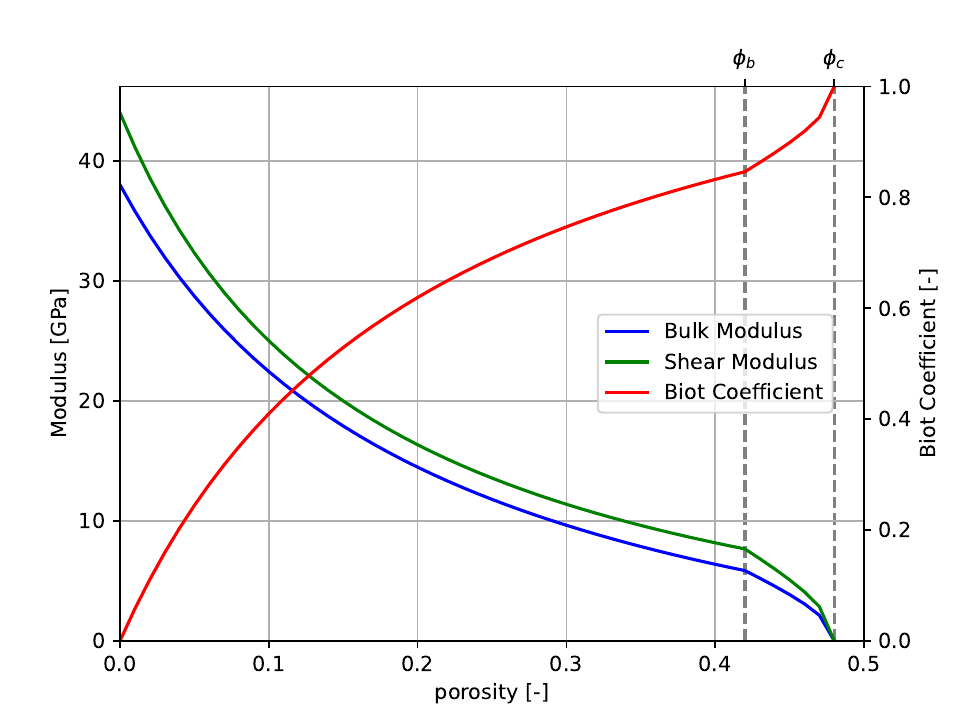}
    \caption{Estimation of rock mechanical properties in response to the cementation}
    \label{fig:constant_cement_plot}
\end{figure}

In contrast to the linear relationship, the constant cement model only relies on material properties and the two porosity parameters ($\phi_c$ and $\phi_b$) that quantify the microstructure of the matrix. The estimated properties with the parameters listed in Tab.~\ref{tab:parameters_cement} are plotted in Fig.~\ref{fig:constant_cement_plot}. By using the relation $\alpha = 1 - \frac{K}{K_s}$, the Biot coefficient can be evaluated simultaneously. Nevertheless, it is crucial to note that this is not the proper way to determine this coefficient and caution must be taken, particularly because the Hashin-Shtriktman bound \cite{Hashin_1963} applied in the last phase could lead to significant deviations between the estimated value at low porosity and the actual value.

This description overlooks the dependence on compressional pressure. Therefore, in the simulation, the variation in moduli is only influenced by changes in porosity due to precipitation (in the current case determined by input parameters) . The deformation itself does not include this change.

\begin{table}[]
    \caption{parameters for constant cement model}
    \centering
    \begin{tabular}{l|l}
    \hline
    bulk modulus of grain $K_s$     &  38 [GPa]\\
    shear modulus of grain $G_s$    & 44 [GPa]\\
    bulk modulus of cement $K_c$ & 98 [GPa] \\
    shear modulus of cement $G_c$ & 35 [GPa] \\
    critical porosity $\phi_c$ & 0.48 [-] \\
    well-sorted porosity $\phi_b$ & 0.42  [-] \\
    \hline
    \end{tabular}
    \label{tab:parameters_cement}
\end{table}

\subsubsection{Hydraulic effects}
Several experimental studies have provided evidence for the effectiveness of carbonate precipitation in reducing the permeability of porous media; similar insights are reported by studies regarding the mathematical description aimed at quantifying this effect. Hommel et al. \cite{Hommel2018} conducted a comprehensive review of the general formulations concerning the relationship between permeability and porosity with a specific emphasis on induced precipitation. Wetzel et al. \cite{Wetzel2020_permeability} numerically investigated that changes in permeability can follow distinct rules contingent upon the governing factor, either reaction kinetics or transport processes. According to this perspective, the alteration in permeability is further categorized into two phases, corresponding to the two phases in the constant cement model. During the initial phase (referred to as the contact cement model), the permeability exhibits behavior analogous to the Kozeny-Carmen equation, Eq.~\eqref{eq: Kozeny-Carmen}, wherein the precipitation primarily impacts the pore throat diameter while preserving the overall shape of the pore throats. In the subsequent phase, a power law model, Eq.~\eqref{eq: permeability power law}, is employed due to the random distribution of precipitation, leading to the possibility of a clogging effect, as previously observed in the work of Weinhardt et al. \cite{Weinhardt2022}.
\begin{equation}
k = \Phi_s^2\frac{\epsilon^3 D^2_p}{180(1-\epsilon)^2}
    \label{eq: Kozeny-Carmen}
\end{equation}

\begin{equation}
    k = b\Phi^n
    \label{eq: permeability power law}
\end{equation}

It is worth noting that precipitation can significantly influence the wetting behavior of fluid phases, particularly in the presence of multiple fluid phases \cite{Hommel2022}. In the van Genuchten model, as the reduction in porosity becomes more pronounced, the fitting parameter $m$ tends to increase, while $\alpha$ tends to decrease. Nevertheless, further investigation is necessary to accurately quantify the impact of precipitation on phase behavior in multi-phase systems. In certain unfavorable scenarios, neglecting the heightened capillary pressure at identical saturation levels could potentially lead to unexpected rock shear failure, particularly as the pore pressure approaches critical limits. This aspect will be discussed upon in the subsequent section. In this paper, for the sake of simplicity, this effect would be roughly presented by adapting the parameters in the capillary pressure - saturation relationship.

\subsection{Rock failure and induced seismic events}
In this section, we dive into the discussion of rock failure, or fault-zone (re-)activation and its associated geomechanical consequences. A commonly employed model in this context is the ubiquitous joint model, using the Mohr-Coulomb criterion with tension cutoff. Since we calculate with an elastic model, we adopt an approach akin to the previous one, wherein the adjustment of the stress tensor is predominantly contingent upon the failure state, with relatively less consideration given to ensuring energy conservation. The primary aim of this methodology is to aptly capture and present the phenomenon of rock failure while preserving the main focus on the injection-introduced failure behavior.

\subsubsection{Residual tensile stress}
\replaced{Rock is commonly recognized as a material with low tensile strength.}{Rock is commonly recognized as a non-tensile-stress bearing material.} In our approach, a maximum tensile stress threshold is established, and when the principal stress exceeds this critical value, a cutoff mechanism is triggered. In essence, this cutoff implies the plastic deformation within the rock material.

During the fluid injection process, the vicinity near the injection area is particularly susceptible to tensile failure. Unlike hydraulic fracturing applications, gas injection projects place considerable emphasis on carefully managing the pressure around the injection well to prevent potential tensile failure. Moreover, this study concentrates on investigating the large-scale impacts of the process. Consequently, the consideration of tensile failure is limited to the initial phase, wherein the initial stress is calculated as a result of the geological formation processes.

\subsubsection{Shear-stress failure / fault reactivation}
Another type of rock failure is the shear-stress failure. When the normal stress acting on the rock matrix is less than the required stress to balance the shear stress, the rock undergoes relative friction, leading to induced \replaced{seismic}{sfeismic} events. To analyze this phenomenon, the stress-drop concept proposed in the work of Beck et al. \cite{Beck-stressdrop} is employed in this study. This concept postulates that failure occurs at the angle where the shear stress surpasses the allowable threshold. Accordingly, a shear-stress drop occurs at the failure angle, predefined in its magnitude, e.g., 1~MPa. The stress drop is associated with dissipation of the energy, though not modelled. If the rock is situated within the fault zone, where the fault angle is known, the stress state at this specific angle is exclusively examined. In such cases, failure occurring within the fault zone is regarded as fault reactivation. 
Algorithm~\ref{algorithm: stress tensor adaption} demonstrates the modification of the incremental rock stress (relative to the previous state) following this assumption. 
It is worthy to note that even the stress tensor is modified after the stress-drop law, the porosity equation \eqref{eq: effctive-porosity} still holds, since the average pressure $\frac{\sigma_{xx} + \sigma_{yy}}{2}$ (the center of Mohr Circle) is unchanged.
\begin{algorithm}[hbtp]
\caption{Stress tensor adaption}
\newsavebox\stressmatrix
\savebox{\stressmatrix}{$\begin{bmatrix}
\sigma_{xx} & \sigma_{xy} - \Delta \tau \\
\sigma_{yx} - \Delta \tau & \sigma_{yy}
\end{bmatrix}$}
\begin{algorithmic}
\State $\boldsymbol{\sigma} = \boldsymbol{\sigma(\Delta u) } + \boldsymbol{\sigma}_{\text{prev}}$
\State \text{//Rotated to principal stress state or by failure angle}: 
\State $\boldsymbol{\sigma} = \boldsymbol{R}^T \boldsymbol{\sigma} \boldsymbol{R}$ \Comment{$\boldsymbol{R}$ as the rotation tensor}
\If{ $\sigma_{xy} > \tau_{\text{max}}$ after Mohr-Coulomb condition}
    \State $\boldsymbol{\sigma}= \usebox{\stressmatrix}$ 
    \Comment{$\Delta \tau$ is stress drop value}
\EndIf
\State{Rotated back to original coordination}: $\boldsymbol{\sigma} = \boldsymbol{R} \boldsymbol{\sigma} \boldsymbol{R}^T$ \\
\Return $\boldsymbol{\sigma} = \boldsymbol{\sigma} - \boldsymbol{\sigma}_{\text{prev}}$
\end{algorithmic}
\label{algorithm: stress tensor adaption}
\end{algorithm}

\subsubsection{Estimating seismic events}
Both shear-stress failure and fault reactivation have significant implications on the local stress tensor, leading to additional deformation in the rock mass. The evaluation of seismic moment in these scenarios can be accomplished using the expression 
\begin{equation}
    M = G \cdot A \cdot S, 
    \label{eq: seismic momentum}
\end{equation}
where $A [m^2]$ represents the area of the crust affected, $S [m]$ corresponds to the average co-seismic deformation, and $G [Pa]$ for the shear modulus at the slip. Again this moment can be translated to the seismic magnitude with the equation
\begin{equation}
    M_w = (\mathrm{log}M_0 - 9.1)/1.5.
\label{eq: sesmic magnitude}
\end{equation}

\subsection{Simulation workflow of geomechanical problems}
The simulation workflow is illustrated in Algorithm~\eqref{algorithm: simulation workflow}. It can be broadly divided into two steps.

In the first step, which may be viewed as a single pre-time step, the initial displacement and stress (Eq.~\eqref{eq: 2p-mass-balance-full}) is computed based on the boundary conditions and the hydrostatic pressure. In many geomechanics simulations, this step is often skipped, assuming a homogeneous lithostatic pressure. However, it is essential to emphasize that such strong simplifications may not accurately reflect the structural composition of the matrix, potentially underestimating the impact of existing structures. 


The second phase, i.e., the main part of the simulation, represents the standard simulation procedure, where we employ the incremental formulation to separate the mechanical effects induced by gas injection. The problem is then solved either in a coupled or decoupled manner, since the hydrostatic pressure is not maintained throughout the simulation. The simulation adheres to a trial pattern, wherein the first step involves calculating the originial displacement, excluding the effect of stress drop ($\Delta \tau = 0$). \replaced{Should any cell undergo shear stress failure, the solution will be adjusted by applying the stress drop value at the failed interface.}{Should any cell undergo shear stress failure, the solution will be to incorporate the stress drop value.} \added{The total slip S is then quantified as the difference between two states.
\begin{equation}
    \mathbf \Delta u = \mathbf u_{t,s} - \mathbf u_t
\end{equation}
\begin{equation}
    \mathbf \Delta u_{\parallel} = (\mathbf \Delta u \cdot \mathbf{d}) \mathbf{d}
\end{equation}
\begin{equation}
    S = \|\mathbf{\Delta u_{{\parallel},l}} - \mathbf{\Delta u_{{\parallel},r}}|,
\end{equation}
where $\Delta u$ is the differece at nodes and $\Delta_\parallel u$ is the displacement along the fault interface with $\mathbf d$ the unit vector of interface. 
}

Upon detecting shear failure, the time step size is sharply reduced to 0.01~s to examine whether the failure propagates to adjacent areas. 
This implicit method facilitates the determination of relative slip resulting from fault reactivation in fault zone. The induced seismic events are subsequently evaluted using the expression mentioned before as post-process step.

\algdef{SE}[SUBALG]{Indent}{EndIndent}{}{\algorithmicend\ }%
\algtext*{Indent}
\algtext*{EndIndent}

\begin{algorithm}
\caption{Simulation workflow}
\label{algorithm: simulation workflow}
\begin{algorithmic}
\State \textbf{Step1}: initial mechanical conditions
\Indent
    \State $solve(\boldsymbol{u}_{t_0}, \boldsymbol{\sigma}_{t_0})$ \Comment{under given hydraulic conditions(e.g. $p_{\alpha,t_0} ,s_{\alpha,t_0}$)}
\EndIndent
\State \textbf{Step2}: time evolution 
\Indent
    \State $solve(\boldsymbol{u}_{t}, p_{\alpha,t}, s_{\alpha,t})$ with $\Delta \tau = 0$ \Comment{no stress drop effect}
    \If{$|\tau| > \tau_{max}$ } \Comment{shear stress failure detected}
        \State $solve (\boldsymbol{u}_{t,s}, p_{\alpha,t,s}, s_{\alpha,t,s})$ with $\Delta \tau$
        \State  $S=S(\boldsymbol{u}_{t,s} - \boldsymbol{u}_{t})$ \Comment{calculate slip}
        \State{$\Delta t = 0.01$}
        \Comment{check for further failures}
    \EndIf
    \State $t_n \rightarrow t_{n+1}$
\EndIndent
\end{algorithmic}
\end{algorithm}


\subsection{Efficient Coding}
In \Dumux, all secondary variables required for the assembly process are encapsulated within the \texttt{volumevariables} class. The coupled problem is solved using a multi-domain approach, allowing the utilization of existing models. However, in the current \Dumux release branch, the caching mechanism for \texttt{volumevariables} in the poro-mechanics model is not supported. Given the incremental formulation (\ref{eq: incremental_formulation}), a previous state is required for flux assembly, necessitating its availability at all times. To address this, we have updated the caching mechanism to store the stress and secondary variables from the previous time step. This enhancement significantly reduces the computational time required for the assembly process.

\section{Verification}
This section aims at demonstrating the successful verification of the introduced model by using various setups and scenarios. The two problems address the standard coupling between flow and geomechanics; they are precisely defined, offering analytical solutions that help verifying the accuracy of the formulations and the implementation of the code.

\subsection{Injection Problem}
The first problem presented in this study is credited to S.~De Simone and J.~Carrera and is referred to as Case~3.2 in De~Simone et al. (2017) \cite{DeSimone2017_benchmark}. The analytical solution for this problem was developed using the conventional flow equation, with adjustments made to account for hydraulic mechanical responses.

\begin{figure}
    \centering
    \includegraphics[width=0.7\columnwidth]{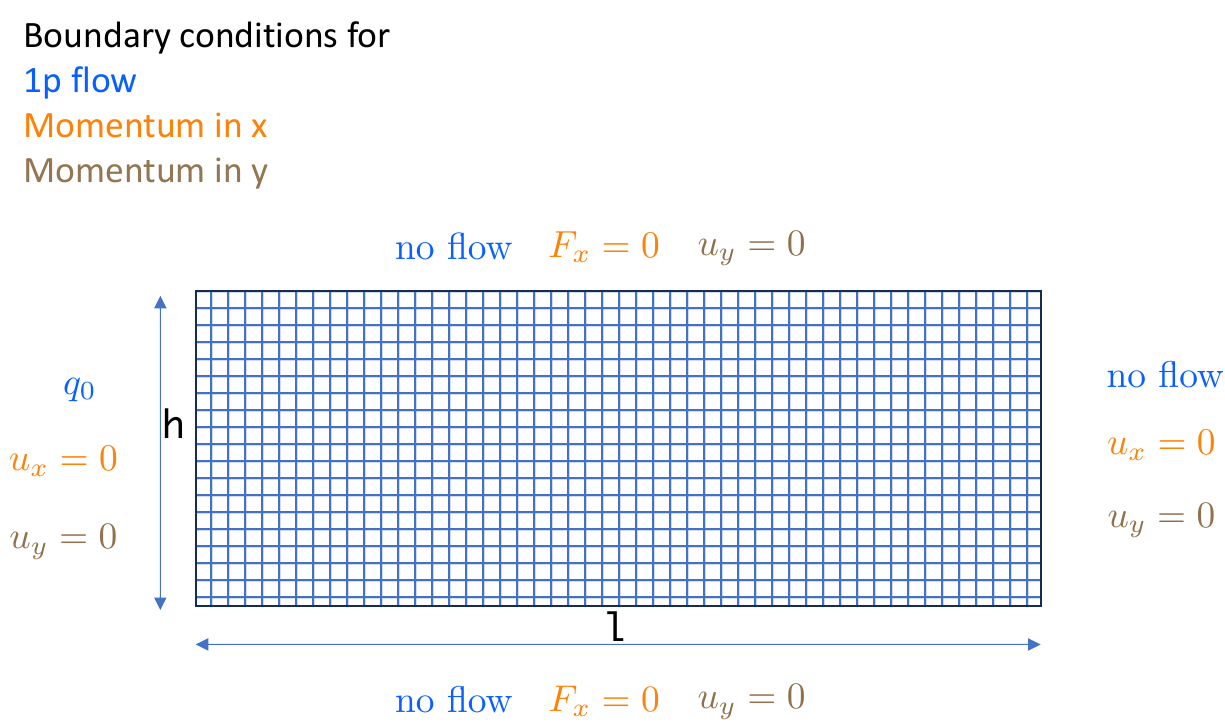}
    \caption{Illustration of the boundary conditions of the domain for the injection problem, i.e., the first verification case.}
    \label{fig:1dbenchmark_bc}
\end{figure}

The problem domain is defined as a finite region with a length of $l$ and a height of $h$. Importantly, the geomechanical deformation within this domain is restricted to the x-direction, i.e., the direction of flow. Water is injected from the left boundary, while no-flow boundary conditions are prescribed elsewhere, see Fig.~\ref{fig:1dbenchmark_bc}. Due to the symmetry in the y-direction, the complexity of the problem can be reduced, simplifying it into a 1D problem.

De Simone and Carrera employ both local and nonlocal storage terms to characterize the hydraulic effect and poro-elastic effect (pertaining to the deformation of aquifers), thereby providing an analytical solution the hydro-mechanical (HM) problem associated with fluid injection in simple geometries. Here, we use the analytical solution as boundary conditions to verify the accurate implementation of equations within the developed \Dumux model.

\begin{figure}
    \centering
    \includegraphics[width=0.7\columnwidth]{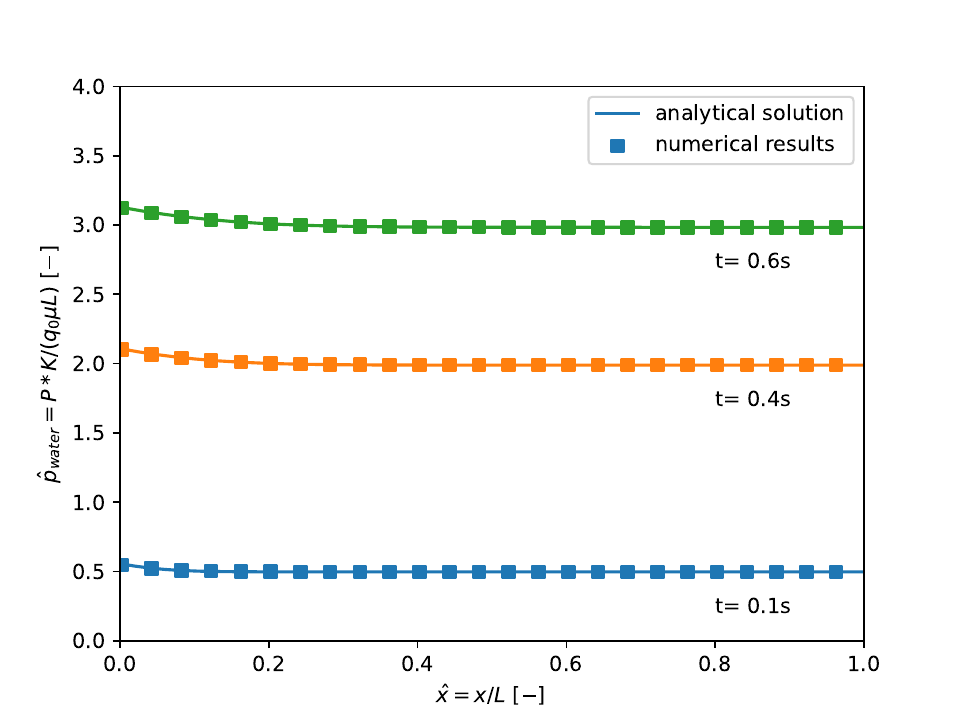}
    \caption{Comparison between the analytical and numerical solutions for the injection problem. $\hat p_{Water}$ is equivalent to $p_d$ in \cite{DeSimone2017_benchmark}.}
    \label{fig:1dbenchmark_results}
\end{figure}

\begin{table}
\centering
\caption{Parameters for 1d benchmark}
\begin{tabular}{l|l}
\hline
compressibilty of water $\beta$ & 5e-10 [1/Pa] \\ 
water viscosity $\mu$ & 1e-3 [Pa$\cdot$ s] \\
porosity $\phi$ & 0.4 [-] \\
permeability $k$ & 1e-10 [$m^2$] \\
Biot coefficient $\alpha$ & 1.0 [-] \\
shear modulus $G$   &  8e6 [Pa]    \\
lame first constant $\lambda$ & 12e6 [Pa] \\
\hline
\end{tabular}
\end{table}

Figure~\ref{fig:1dbenchmark_results} demonstrates the excellent agreement between the analytical and the numerical solutions when employing the incremental formulation and the previously mentioned porosity alteration relationship (Eq.~\eqref{eq: effctive-porosity}). As the injection process progresses, an over-pressure phenomenon becomes evident throughout the domain. Notably, the over-pressure becomes more significant as we approach the injection boundary.

\subsection{Mandel's Problem}
Mandel's problem is a well-studied case in poro-elasticity, see\cite{Keilegavlen_2020, phi2007coupling}. It was initially introduced in the paper by Mandel \cite{man1953geotechnique}, and later extended to anisotropic conditions by Abousleiman et al. \cite{abo1996mandel}. The problem entails investigating the deformation of a water-saturated 2D matrix subjected to constant-force boundary conditions and the drainage of water. The well-known Mandel-Cryer effect, a hydromechanical response, demonstrates the non-monotonic change in water pressure during the drainage process. Due to the intriguing nature of this hydromechanic response, Mandel's problem is a valuable choice for evaluating and examining the model.

\begin{figure}
    \centering
    \includegraphics[scale=0.7]{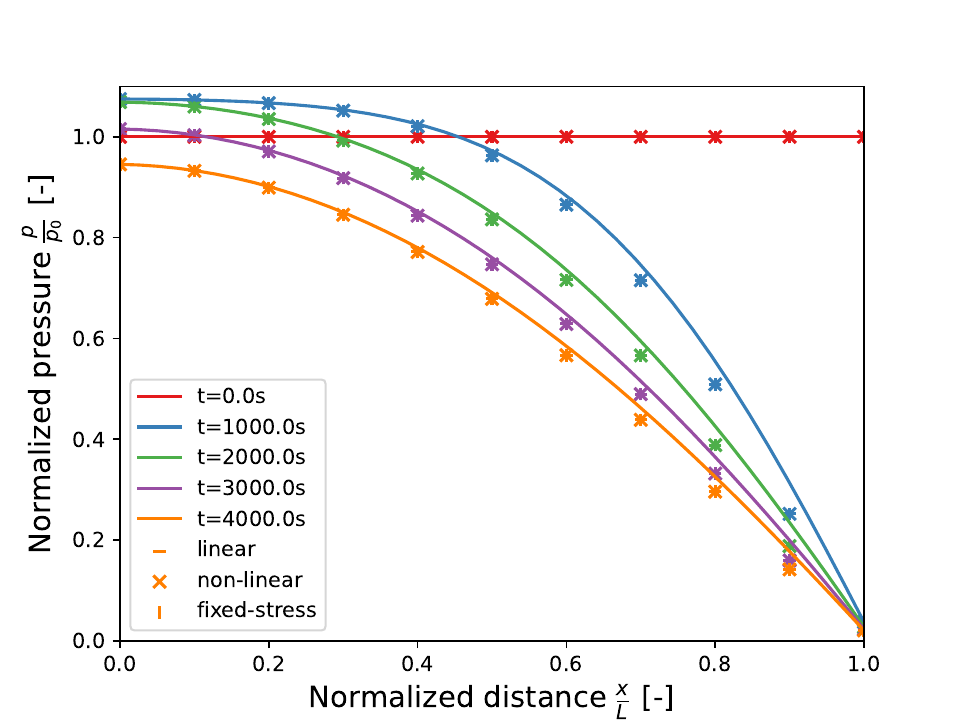}
    \caption{Verification of Mandel's problem}
    \label{fig:verifactation of mandel}
\end{figure}

For the verification, only the isotropic case is considered. The analytical solution is simplified from the one in \cite{abo1996mandel}. Exploring the symmetry of the domain and the boundary condition, only one fourth of the domain is simulated. 

Fig.~\ref{fig:verifactation of mandel} demonstrates that our model successfully reproduces the evolution of pressure. In the plot, we compared the results obtained using the linear formulation Eq.\eqref{eq: linear storage term} with those from our non-linear formulation Eq.\eqref{eq: 2p-mass-balance-full}.

Furthermore, we examined the decoupled scheme (fixed-stress splitting), which yielded converged values identical to those obtained by the fully coupled formulations. Yet, this comparison verifies the accuracy and reliability in incremental formulation by applying the appropriate incremental boundary conditions. 

Lastly, we examined the computational improvements attributable to the \Dumux advancements in cache handling. Under identical conditions, we observed that the assembly time per Newton iteration with caching was reduced from 0.25 seconds to 0.125 seconds, corresponding to an approximate $50\%$ reduction. Notably, the solution time for the linear solver remained unchanged.

\section{Comparison between simulated and experimental data of an EICP-Treated sand column}
\label{sec:discussion icp column}

The constant cement model was originally proposed to analyze rock formation and the mechanical role of cement \cite{Avseth2010}. We propose utilizing this model in our study to predict the evolving mechanical properties of the matrix at varying porosity or cement content. Apparently, there is a need to test its applicability and to get a better understanding of its limitations.

Due to the lack of well-documented and well-controlled experimental data on real core samples undergoing biomineralization, we have employed the experimental data for artificially produced sand columns from the study conducted by Yasuhara et al. \cite{yasuhara2012experiments}. In their study, the authors created columns using EICP methods, wherein they initially mixed Toyoura Sand with the urease (enzyme) before injecting cementation solutions ($\text{CaCl}_2$ and urea) from the bottom into the sand column (5~cm in diameter and 10~cm in height) under fully saturated and compressed conditions. Following several injections, the cementation process bonded the sand particles together to form samples for further testing. Subsequently, the column was dried out, and an acid leaching method was employed to measure the weight and volume occupied by the cement. Additionally, the authors conducted Unconfined Compression Strength (UCS) tests on the samples for mechanical properties. A group of experiments were conducted, varying in the mixing ratio of sand and urease, the concentration of the cement solution, and numbers of injection. The initial sand samples had a porosity of $\phi_{ini} =0.44$ and the acid leaching revealed that calcite cement accounted for approximately 0.01 to 0.04 of the column's volume. The UCS tests showed a range of the secant elastic modulus at $50\%$ of the peak strength $E_{50}$ values from 53.5 to 160 MPa. Additionally, the results of X-ray diffractometry (XRD) confirmed that quartz (grain material) and calcite (cement material) are the main components of the treated sand columns. 

\replaced{The conducted experiment aligns with the requirements of the constant cement model, as the parameters needed for this model can be directly deduced from the experimental data, including the UCS and porosity data by}{The conducted experiment coincides with the cementation theory, and therefore we chose the UCS data and the porosity data by } 
\cite{yasuhara2012experiments}
 as the reference for the predictions provided by the constant cement model. Given that our focus in this section is on the mechanical aspect, we exclusively simulate the UCS tests of dry samples as a purely mechanical process. \added{The setup and balance equations are listed in Appendix C.} We use a 2D domain (2.5 cm in radius and 10 cm in height) to represent the half of the column cross section, under the strong assumption that the distribution of precipitation within the column samples is axially symmetric. The compressive force is represented as the displacement of rigid surface on the top, while the average normal stress on the top is considered as the value obtained from the UCS test. The load velocity of strain is set at $0.045\% \frac{cm}{cm \cdot s}$, with the simulation ending at a total strain of $0.45\%$. Elastic moduli of quartz ($K_s= 38~GPa, G_s= 44~GPa$) and calcite ($K_c= 63~GPa, G_c= 31~GPa$), critical porosity $\phi_c = \phi_{ini} = 0.44$ are given as input parameters to the constant cement model. However, there is no direct data available for estimating the important value of the well-sorted porosity $\phi_b$, and thus, we adopt a range from 0.40 to 0.44, as suggested in other work \cite{Criticalporosity2001}. Furthermore, we consider the heterogeneity of precipitation distribution, as the chemical processes maybe influenced by flow conditions within the column. Due to limited data on local porosity, we assume a linear porosity profile (0.39 at the bottom and 0.43 at the top) to investigate the role that the porosity profile plays.

\begin{figure}[hbtp]
    \centering
    \includegraphics[scale=0.7]{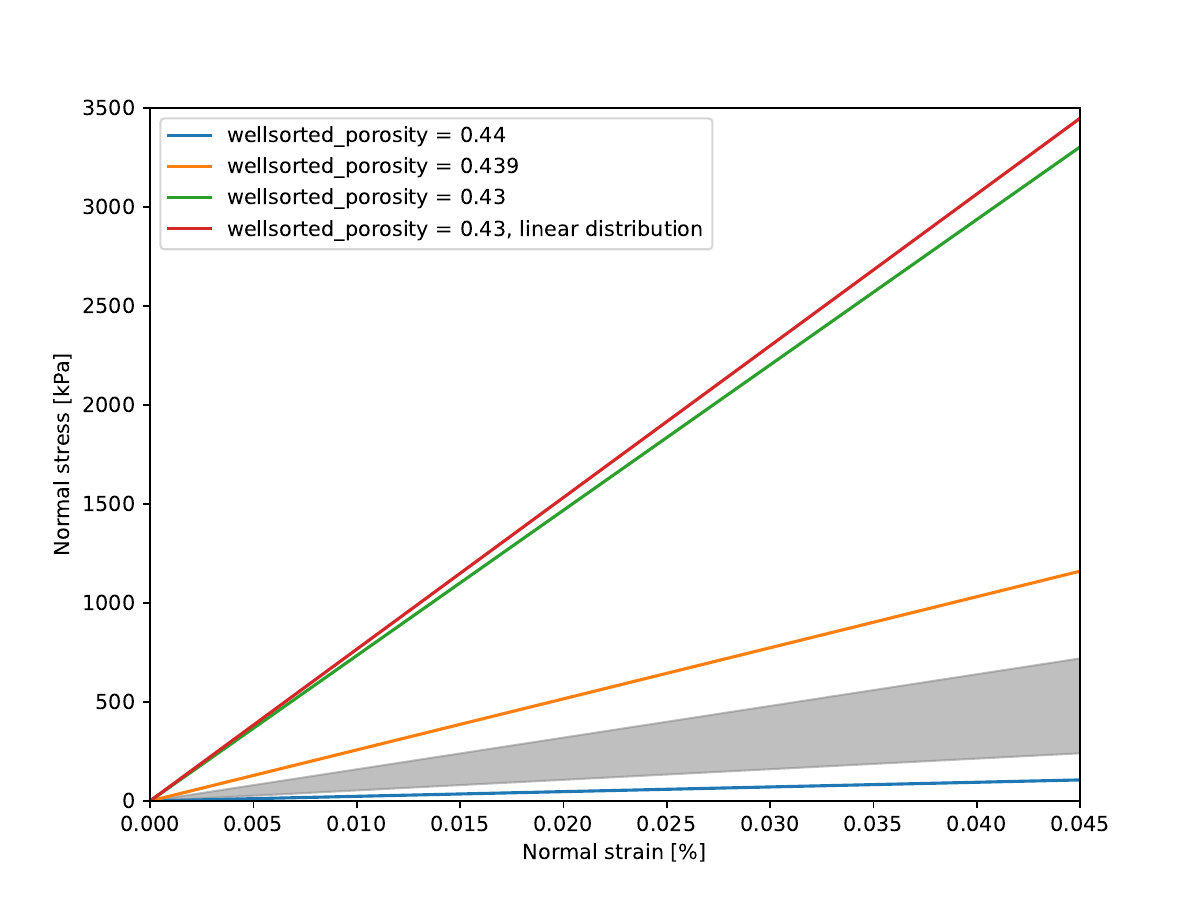}
    \caption{Fictive Unconfined Compression Test Protocol. The gray area represents the measurement in experiments of \cite{yasuhara2012experiments}}.
    \label{fig:ucs_result}
\end{figure}

 Figure \ref{fig:ucs_result} depicts a hypothetical protocol of UCS tests. The experiments show a range of $E_{50}$ values from 53.5 to 160 MPa, while the simulated values range from 23 to 734 MPa. The results indicate that the constant cement model is highly sensitive to the well-sorted porosity. Experimental measurements indicate the well-sorted porosity to fall within a narrow range from 0.439 to 0.44, suggesting potential difference in sorting between natural sands of greater non-uniformity and those used in the Yasuhara et al. experiments. Nonetheless, this finding supports the notion that some precipitation occurs at the contact area, enhancing the compressibility of the entire column. Lastly, the results indicate further that an uneven distribution, in conjunction with the non-linearity in the cement model, could lead to a higher effective $E_{50}$ strength.
 
In summary, the constant cement model demonstrates applicability in predicting strength of EICP-treated porous media, while the  range of predictions for these particular reference experiments is large. We should note that the model is quite simple, requiring only few details; on the other hand, knowledge of the required data is also limited. We suppose that a particular difficulty in transferring insights from these reference data to realistic rocks is the uniformity of grain sizes and heterogeneities, which are likely higher in realistic cases, and, therefore, the sensitivity of the predicted strength to the well-sorted porosity is smaller. In any case, the observed discrepancies highlight the need for further research on detecting and analyzing the role of local precipitation and parameterization.

\section{Effects of biomineralization on geomechanics in a reservoir injection scenario}
Building upon the validated and tested model, as detailed in the preceding sections, we have developed a simulation scenario to examine the practical implications of cementation processes in underground gas storage systems. The scenario is inspired by the CO$_2$ storage framework proposed by Alberto et al.\cite{maz2012induced}, which investigates the potential for fault reactivation and induced seismicity during underground CO$_2$ injection. In this study, we extend the framework to include the effects of biomineralization, with a continued emphasis on fault reactivation, while explicitly excluding rock failure within the matrix. Given the significant uncertainty associated with strength predictions, as highlighted in the previous section, this scenario is intended as a case study to discuss the implications of biomineralization, utilizing results that are believed to represent reasonable orders of magnitude.

\subsection{Scenario definition}
\begin{figure}
     \centering
     \includegraphics[width=0.8\linewidth]{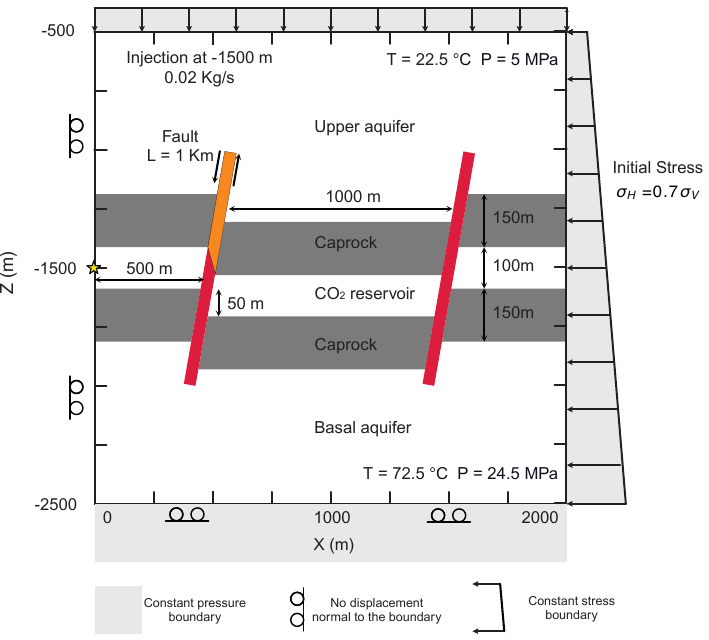}
     \caption{Illustration of the setup and boundary conditions for the scenario.}
     \label{fig:showcase-setup}
 \end{figure}

As shown in Figure \ref{fig:showcase-setup}, the proposed two-dimensional domain measures 2 km by 2 km, with a depth range from 500 m to 2500 m. The storage layer, which is 100 meters thick, is bounded above and below by less permeable rock layers, each 150 meters thick. These bounding layers are themselves surrounded by two other formations that have permeability comparable to the storage layer. To investigate the latent effects of biomineralization on the original geological condition, we modified the fault zone configuration by introducing a \replaced{second}{secondary} fault zone and adjusting the offset of two faults. Within the subsurface, two fault zones are present at a distance of 1 km apart, each with a dip angle of 80° and a width of 10 meters, intersecting the storage layer. For this showcase scenario, we assume that the left fault zone, which is more permeable, is detectable and indicates the need for biomineralization, whereas the right fault zone is less permeable and no treatment occurs there.

In practical terms, biomineralization will only seal the upper section of the left fault zone, as indicated by the orange marking in the figure. These fault zones initially exhibit a shear displacement offset (throw) of 50 meters, resulting in a hydraulically confined reservoir. Specifically, the left fault zone has a positive throw of 50 meters, while the right fault zone exhibits a negative throw of 50 meters. Carbon dioxide (\ce{CO2}) is injected from the left boundary at a constant rate of $0.0004~\text{kg}/(\text{m}^2\cdot \text{s})$, corresponding to $630.72~\text{t/year/m}$ for a site extending 100 m in length. The injection process is terminated upon detecting fault reactivation in any of the fault zones, with the simulation concluding in all cases after 100 days.

Three distinct test cases are defined to evaluate the effects of different treatment strategies on fault zones. Case A serves as the baseline scenario, in which neither of the fault zones receives any treatment, allowing for natural conditions to be observed. In Case B, a targeted partial zone undergoes slight sealing through induced biomineralization, representing a moderate intervention. Case C involves an intensive treatment of the fault zone, which significantly reduces porosity. This scenario examines the impact of extensive biomineralization on the structural properties of the fault zone. 

The hydraulic and mechanical parameters are presented in detail in Table~\ref{tab:material_parameters_input}. For the left fault zone, the parameters are divided into three cases, with the lower part of the zone remaining unchanged across all simulations, using the same values as those listed for Case A. The initial porosities correspond to the general values used in the original setup. For moderate and intensive treatments, reductions of $8\%$ and $10\%$ in porosity, respectively, are considered. \added{Referencing to experimental observations in \cite{Weinhardt2022}}, the permeability of the treated fault zones is calculated using the power law described in Eq.\eqref{eq: permeability power law}, with $n=3$, leading to a reduction by an order of magnitude. The parameters for the Van Genuchten Model are estimated based on experimental trends observed by \cite{Hommel2022}, with general factors of 0.33 and 0.2 applied for $V_{\alpha}$ and $V_n$, respectively in both cases, as the actual porosity approaches. 

The mechanical part employs the constant cement model, utilizing the parameters outlined in Table~\ref{tab:parameters_cement}. The bulk and shear moduli are directly assigned for specific geological zones, rather than being treated as fitted parameters, to ensure consistency with other simulations. For the fault zones, the critical porosities are assumed to be slightly higher than the initial porosity, reflecting the potential for poorly cemented structures within the geological formations. Furthermore, the well-sorted porosity is set to be $3.5\%$ less than the critical value.


\begin{sidewaystable}[hbtp]
    \centering
    \caption{Summary of Material Properties by Geological Zone}
    \label{tab:material_parameters_input}
    \begin{tabular}{llcccccccc}
        \toprule
        \multirow{2}{*}{} & \multirow{2}{*}{} & \multirow{2}{*}{Caprock} & \multirow{2}{*}{Reservoir} & \multirow{2}{*}{\begin{tabular}[c]{@{}c@{}}Upper \\ Aquifer\end{tabular}} & \multirow{2}{*}{\begin{tabular}[c]{@{}c@{}}Basal\\ Aquifer\end{tabular}} & \multirow{2}{*}{\begin{tabular}[c]{@{}c@{}}Right\\ Fault\end{tabular}} & \multicolumn{3}{c}{Left Fault} \\ \cmidrule(lr){8-10} 
        & &  &  &  &  &  & A & B & C \\ 
        \midrule
        & {Porosity [-]} & 0.01 & 0.1 & 0.1 & 0.01 & 0.1 & 0.15 & 0.07 & 0.05 \\ 
        \midrule
        \multirow{3}{*}{\rotatebox[origin=c]{90}{Hydraulic}} & {Van Genuchten $\alpha$ [1/Pa]} & $5.025 \times 10^{-5}$ & $5.025 \times 10^{-4}$ & $5.025 \times 10^{-4}$ & $5.025 \times 10^{-4}$ & $5.025 \times 10^{-4}$ & $5.025 \times 10^{-4}$ &  $1.675 \times 10^{-4}$  & $1.675 \times 10^{-4}$ \\ \cmidrule(lr){2-10} 
        & {Van Genuchten N [-]} & 1.842 & 1.842 & 1.842 & 1.842 & 1.842 & 1.842 &  0.368 & 0.368 \\ \cmidrule(lr){2-10} 
        & {Permeability [$\mathrm{m}^2$]} & $1 \times 10^{-19}$ & $1 \times 10^{-13}$ & $1 \times 10^{-14}$ & $1.9 \times 10^{-16}$ & $1 \times 10^{-15}$ & $1.9 \times 10^{-13}$ & $1.9 \times 10^{-14}$ & $7 \times 10^{-15}$ \\ 
        \midrule
        \multirow{5}{*}{\rotatebox[origin=c]{90}{Mechanical}} & {Biot Coefficient [-]} & 0.242 & 0.778 & 0.778 & 0.778 & - & - & - & - \\ \cmidrule(lr){2-10} 
        & {Bulk Modulus [MPa]} & $34.4$ & $6.0$ & $6.0 $ & $6.0 $ & - & - & - & - \\ \cmidrule(lr){2-10} 
        & {Shear Modulus [MPa]} & $18.7$ & $8.0 $ & $8.0$ & $8.0$ & - & - & - & - \\ \cmidrule(lr){2-10} 
        & {Critical Porosity [-]} & - & - & - & - & 0.105 & 0.155 & 0.155 & 0.155 \\ \cmidrule(lr){2-10} 
        & {Well Sorted Porosity [-]} & - & - & - & - & 0.07 & 0.12 & 0.12 & 0.12 \\ 
        \bottomrule
    \end{tabular}
\end{sidewaystable}

\subsection{Results}

\begin{figure}[hbtp!]
    \centering
    \includegraphics[width=0.6\linewidth]{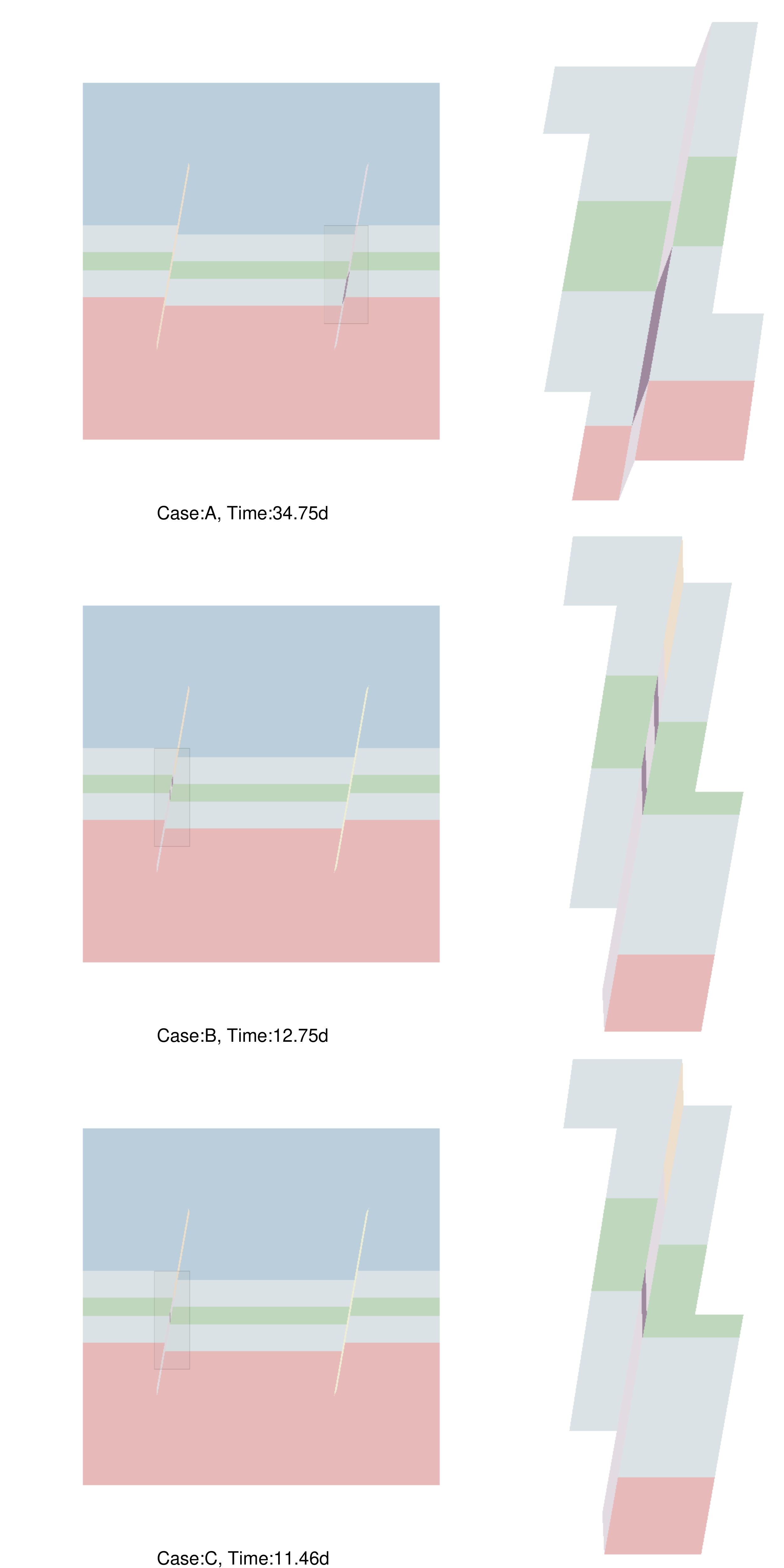}
    \caption{Visualization of cells experiencing shear stress failure during the initial failure event. The colors represent different zone types, with the darkest color indicating the failed zones. The time indicated in the figure corresponds to the moment when the first failure event occurs.}
    \label{fig:failure_cell_positions}
\end{figure}

In three instances, a single injection-induced shear failure event was observed at 34.75~days, 12.75~days, and 11.46~days, respectively. These observations indicate that a less permeable and more robustly sealed zone leads to faster failure events. In addition to influencing the timing of occurrence, the biomineralization process also alters the failure position. Figure~\ref{fig:failure_cell_positions} illustrates the locations of the failed cells in each scenario. In the absence of biomineralization, failure occurred at the intersection of the right, less permeable zone and the lower rock layer. With biomineralization applied, primary failures shifted to the left fault zone. All failures occurred then in the untreated left fault zone and remained within the reservoir layer. In Case B, two failure cells were detected at the connection points to the neighboring rock layer, whereas in the stronger cemented Case C, only the lower cell experienced shear stress failure. \added{Here, we focus on the ICP effect within the fault zone; additional results concerning injection-related pressure and stress changes are provided in Appendix D.}

\begin{figure}[hbtp!]
    \centering
    \includegraphics[width=\linewidth]{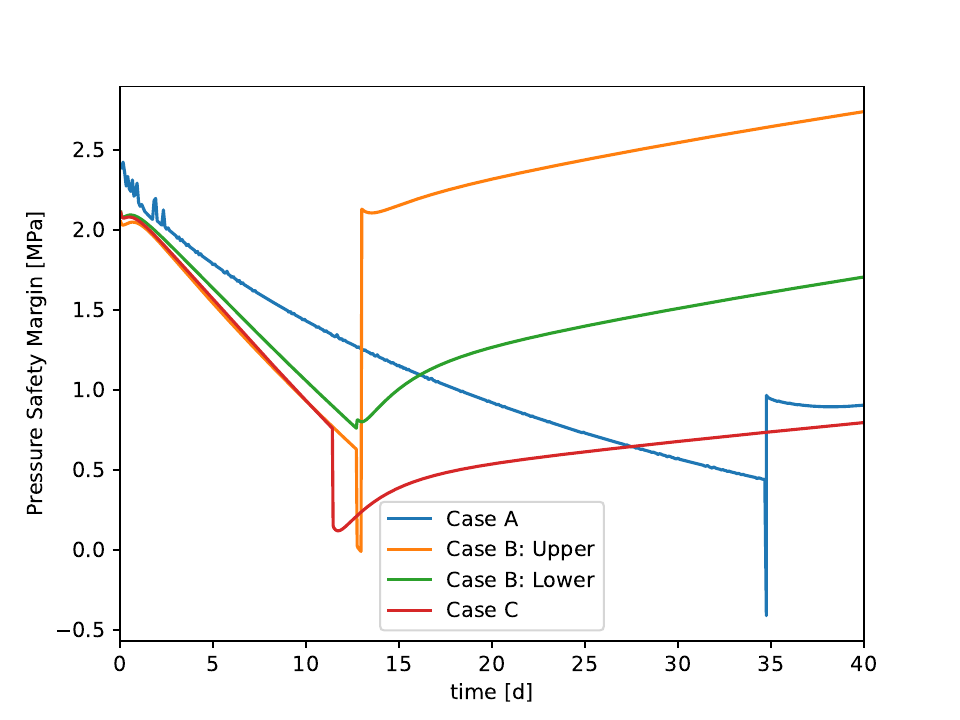}
    \caption{Time-dependent safety margin of failed cells across different cases.}
    \label{fig:pressure_safety_margin}
\end{figure}

We use the concept of "pressure safety margin" to indicate the maximum allowed pressure change necessary to prevent triggering shear stress failure.\added{
\begin{equation}
    \tau' = (\sigma' -\alpha P_{Pore} - P_{Safety}) \mathrm{tan}\phi +c,
\end{equation}
where $\tau'$ and $\sigma'$ represent the shear stress and normal stress along the fault direction, respectively.The parameters  $\phi$ and $c$ denote the friction angle and cohesion coefficient, as defined in the Coulomb–Mohr theory.} This margin can also be interpreted as the distance from the failure curve line to the Mohr circle. Notably, as the safety margin approaches zero, the system becomes highly sensitive to shear failure. Figure~\ref{fig:pressure_safety_margin} shows the change in the pressure safety margin of selected cells over time. In Case A, we selected the cell with the most significant stress change as the representative one. The two cells in Case B and the single cell in Case C are also plotted. The timing of the dramatic drop in the safety margin corresponds to occurrences of shear stress failure, although the drop value (ranging from 0.1~MPa to 2.2~MPa) does not match the shear stress drop value (1.0~MPa). All curves confirm that the induced stress drop enhances safety subsequently after the shear stress drop. It is also apparent that the safety margin value drops at a similar rate in the treated cases (B and C) during the injection period.

\begin{figure}
    \centering
    \includegraphics[width=0.8\linewidth]{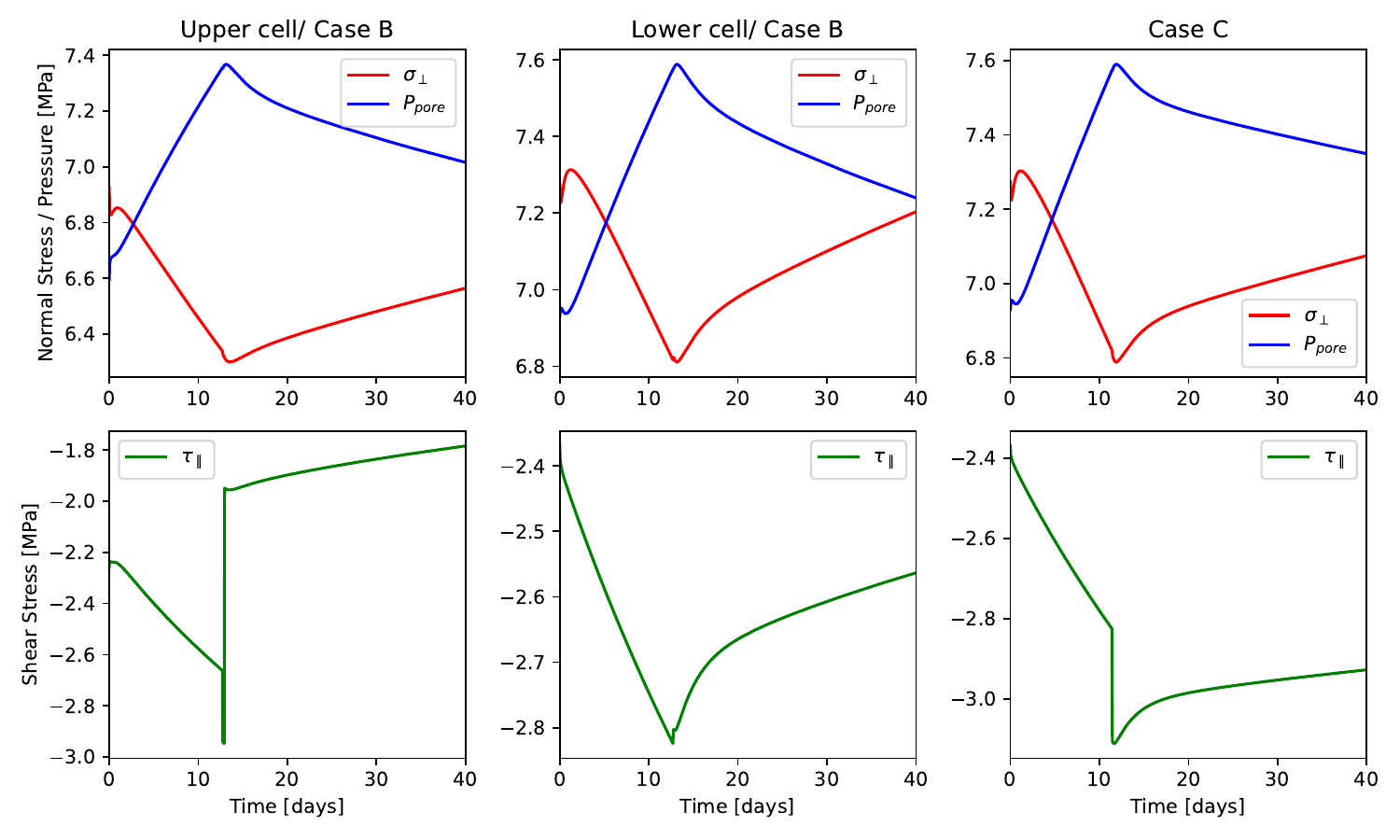}
    \caption{Time evolution of normal compressive stress, shear stress on the fault plane, and the effective pore pressure change for the failed cells in Cases B and C.}
    \label{fig:stress_evolution}                               
\end{figure}

A further analysis of the stress tensor indicates that the principal stress state is very close to the physical configuration, with an angle of less than 2 degrees. The local rotated stress state along fault zone is also similar to the principal one, as the dip angle of the fault zone is nearly parallel to the axis. Figure~\ref{fig:stress_evolution} provides more details about the temporal evolution of the stress and pressure states in the failed cells. $\sigma_{\tau}$ represents the normal compressive stress (positive for compression) on the fault plane, while $\tau$ denotes the local shear stress on the fault zone. The injection operation results in an increase in pore pressure (around 1~MPa) and shear stress (around 0.4~MPa), as well as a decrease in normal compressive stress (around 0.5~MPa). Additionally, the shear stress change in the lower cell in Case B shows a significant difference in stress drop value, while the other cell exhibits only a slight change.

\begin{figure}[hbtp!]
    \centering
    \includegraphics[scale=0.6]{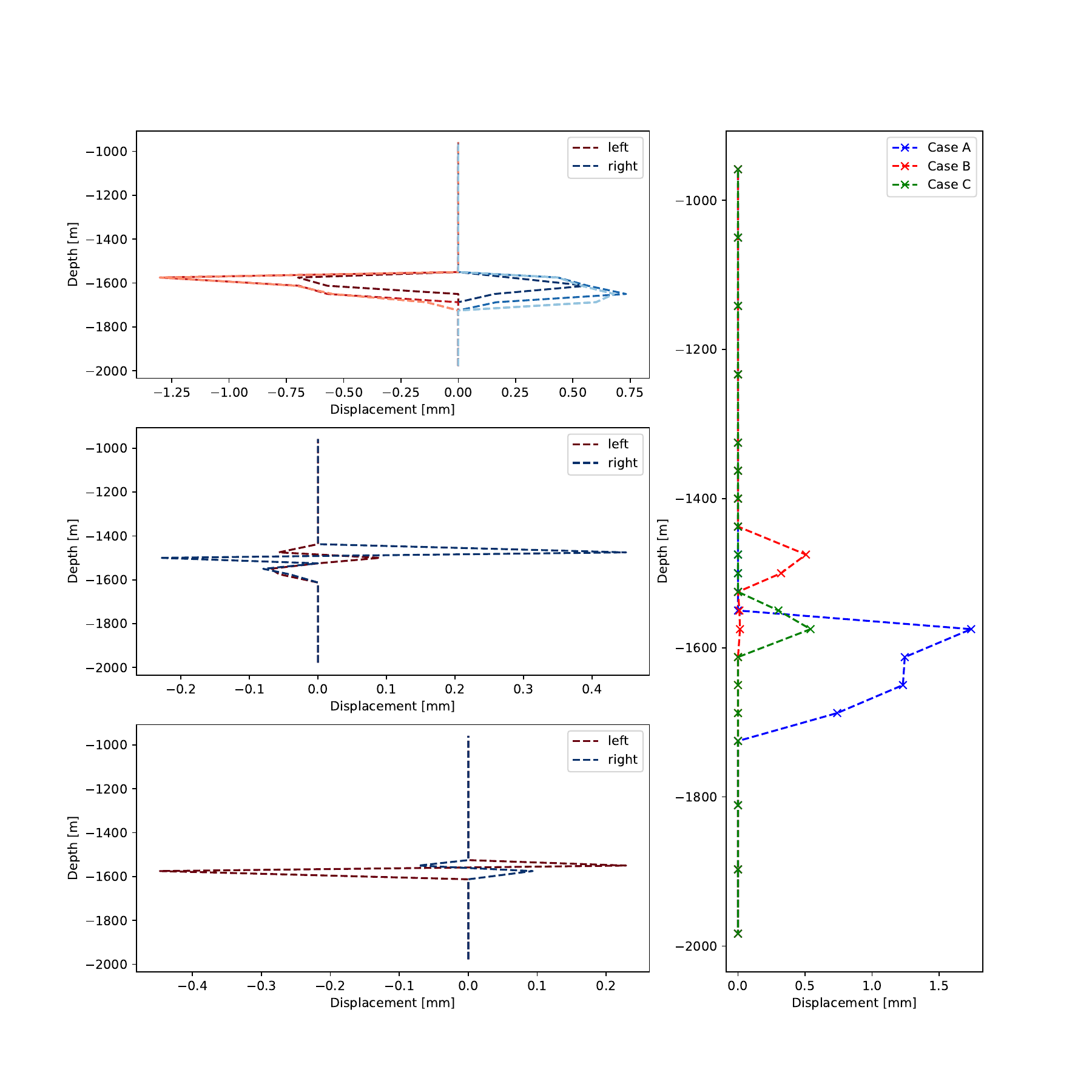}
    \caption{Relative displacement (left) and total slip (right) along the fault zone. The panels on the left, from top to bottom, correspond to Cases A, B, and C. In the left panels, lighter colors indicate events occurring 0.01 seconds after the previous event.}
    \label{fig:relative-slip-both-scenarios}
\end{figure}

The difference in displacement and slip of the fault zone in each case is depicted in Figure~\ref{fig:relative-slip-both-scenarios}.  In Case~A, significant deformation occurs within 0.03~s, where shear failure in the initially failed cell also triggers adjacent cells, resulting in substantial slip at the end. The displacement in the other two cases is of a similar magnitude, where we observe opposite slip directions and pronounced asymmetrical displacement. Assuming a field length of 100~m, the slips correspond to seismic events of magnitudes 1.16, 0.63, and 0.64. It is evident that mineralization alters the positions where seismic events are triggered but effectively reduces seismic energy since the occurrence is earlier. However, it is important to note that an earlier occurrence of failure does not necessarily indicate a lower intensity.

\section{Discussion and Conclusions}
Sealing flow paths to prevent leakage is a primary target of biomineralization and affects the hydraulic properties, in particular porosity and permeability.
Beyond that, the results of the showcased scenarios highlight the practical potential impact of biomineralization on geomechanical properties and events in underground gas storage. Gas injection reduces compressive stress due to the effective stress effect and simultaneously induces additional shear stress within the domain to maintain elastic deformation of the subsurface. This result is significant, as it aligns with reported events observed in real-world field operations. However, this effect is subject to variations in geological formations and boundary conditions, which are not covered in the showcase.

Within the proposed scenarios, a locally sealed fault zone slightly affects the flow paths for injected gas. However, the rapid buildup of pressure, due to fluid compression and the pressure gradient required for flow, depletes the pressure safety margin. This depletion may explain why the failure is shifted. In addition to the hydraulic effects, the process also increases the stiffness of the fault zone. A higher elastic modulus allows for smaller deformations, resulting in less additional shear stress, which enhances the safety of the matrix. However, it also decreases the homogeneity of the domain, causing weaker sections to undergo larger deformations and potentially fail, as observed in this case.

A different extent of the biomineralization process exhibits similar changes in effective pore pressure and shear stress. This similarity can be attributed to the flow field not being strongly affected. However, the displacement of the cells shows a markedly different pattern of deformation, which can be attributed to the occurrence of failure at different positions. In Case B, the upper failed cell undergoes most of the slip, providing the lower cell with additional freedom. In contrast, in Case C, the failed cell is located in the lower half of the domain and lacks adjacent cells to absorb the additional energy, resulting in deformation in a different direction. 
As shown in Fig.~\ref{fig:constant_cement_plot}, the variation in properties demonstrates that a stiffer sealed fault zone results in increased plastic deformation in the surrounding weaker regions. This observation is further corroborated by the corresponding seismic energy levels.

Focusing on the energy levels, it is apparent that biomineralization results in much smaller seismic activities compared to the cases without biomineralization. In contrast to the aftershocks reported in \cite{Keranen2014}, no post-injection seismic events were detected. This can be attributed to the setup and characteristics of the proposed system, where failures occur much earlier and injection stops immediately after the occurrence. However, given the rather low intensity of the activity, an instantaneous response is unfeasible in practical scenarios, leading to a substantial buildup of pore pressure. This lack of buildup also contributes to the absence of post-injection events in the modelled scenarios.

\replaced{The presented study demonstrates the potential effects of biomineralization in a gas-storage scenario with fault zones. However, the results are highly dependent on the specific geometric setup and model parameters, making them site-specific and difficult to generalize. While the model could be applied to CO$_2$ mineralization scenarios, this was beyond the scope of the study. The constant cement model used relies on several parameters that should be validated with real-world data, and the study did not account for local effects or heterogeneity within the geological zone. From an engineering perspective, many operational possibilities remain unexplored.
Despite the promising results, several limitations should be noted. The constant cement model is heuristic and simplifies the mechanical properties of rock grains, cements, and porosity parameters, which can vary widely in reality. The permeability-porosity relationships depend on experimental data but do not fully account for the spatial distribution of mineral precipitation, indicating the need for more detailed subscale analyses. Moreover, the model assumes small deformations and linear elasticity, limiting its applicability to more complex or non-linear geological scenarios. Future work should refine these assumptions, integrate more comprehensive experimental data, and extend the framework to include more sophisticated mechanical and flow behavior to improve model robustness and applicability.}{The presented showcase demonstrates and illustrates the potential effects of biomineralization in a gas-storage scenario with fault zones; however, the results are strongly dependent on the specific geometric setup and the model parameters, i.e., they are site-specific and, thus, difficult to generalize. The constant cement model used here relies on several parameters that should be validated with real-world data and experience. Furthermore, the study did not consider local effects and heterogeneity within the geological zone. From an engineering perspective, many operational possibilities have yet to be investigated.}

At the end, we can briefly summarize this study as follows:
\begin{itemize}
\item We further developed and updated the geomechanical model of the open-source simulation platform \Dumux to its latest version.
This update preserves the stress drop concept from previous work to describe shear stress failure, while employing an incremental formulation as an interface to facilitate future improvements and enhance computational efficiency.
The model is tested and successfully verified against benchmarks from the literature.

\item We developed and applied a method to integrate changes in the hydraulic and mechanical properties of porous media under biomineralization. These effects are linked to pore space alteration, and a preliminary comparison with models and experimental data from the literature demonstrates the potential of this approach.

\item We further applied the proposed model to investigate the potential effects of biomineralization on the field scale of underground gas storage scenarios. Although the showcased scenario relies on a significant abstraction of reality, our study demonstrates that this process can alter system behavior and reduce damage to the fault zone. Additionally, the simulations highlight the intricate effects of coupled processes (sealing effect vs. seismic energy) inherent to the showcased scenario, which must be accounted for in real-world decision-making procedures. A well-sealed fracture is desired for preventing leakage, while strong permeability reduction and corresponding strong stiffness increase have effects on potential seismic events and a sound understanding of these relations needs to be further developed for the engineering practice. 
\end{itemize}

\textbf{Data Availability}
The code is available as a dataset hosted at the data repository of the University of Stuttgart (DaRUS)\cite{darus-4457_2024} and as a DuMux-pub module https://git.iws.uni-stuttgart. de/dumux-pub/wang2024a, accessed on 29 August 2024.

\textbf{Acknowledgement}
We thank the Deutsche Forschungsgemeinschaft (DFG, German Research Foundation) for supporting this work by funding SFB 1313, Project Number 327154368.

\newpage
\bibliography{main}

\newpage
\begin{appendices}
\section{Effective Porosity}
\begin{align}
    \Delta \phi =& - \frac{\alpha - \phi}{K} \Delta p  \nonumber \\
    \Delta \phi =& - (\frac{1-\phi}{K} - \frac{1}{K_s}) (-K \epsilon +(\alpha -1 ) p_{eff}) \nonumber \\
    \Delta \phi =& (1-\phi - \frac{K}{K_s}) \epsilon + (\frac{1-\phi}{K} - \frac{1}{K_s}) \frac{K}{K_s} p_{eff} \nonumber \\
    \Delta \phi =& \alpha \epsilon - \phi \epsilon + (\frac{1}{K_s} - \frac{\phi}{K_s} - \frac{K}{K_s^2} ) p_{eff} \nonumber \\
    \Delta \phi =& = \alpha \epsilon + (\frac{1}{K_f} + \frac{1}{K_s} - \frac{\phi}{K_s} - \frac{K}{K_s^2}) p_{eff} \nonumber\\
    \Delta \phi =& \alpha \epsilon + \frac{1}{M} p_{eff}
    \label{eq: linear porosity law},
\end{align}
here $\epsilon = - \frac{\Delta V}{V} = - \frac{\Delta V_p}{V_p} =  \frac{p_{eff}}{K_f}$, with $K_f$ as the effective bulk modulus of the fluids (pore pressure in response to pore volume change). If only singe phase fluid is present, the formulation is identical to the one in \cite{Coussy2003}. \\

\section{Constant Cement Model}
Phase 1: $\phi_c \rightarrow \phi_b $

\begin{equation}
K = \frac{n(1 - \phi_c) M_c S_n}{6},
\end{equation}

\begin{equation}
G_ = \frac{3 K_{\text{dry}}}{5} + \frac{3n(1 - \phi_c) G_c S_\tau}{20},
\end{equation}

where the coefficients $S_n$ and $S_\tau$ are determined by specific equations that take into account the elastic moduli of the grain material and cement material, critical porosity, as well as various statistical parameters.

\begin{equation}
S_n = A_n(\Lambda_n) \alpha^2 + B_n(\Lambda_n) \alpha + C_n(\Lambda_n)
\end{equation}

\begin{equation}
A_n(\Lambda_n) = -0.024153 \Lambda_n^{-1.3646}
\end{equation}

\begin{equation}
B_n(\Lambda_n) = 0.20405 \Lambda_n^{-0.89008}
\end{equation}

\begin{equation}
C_n(\Lambda_n) = 0.00024649 \Lambda_n^{-1.9864}
\end{equation}

\begin{equation}
S_\tau = A_\tau(\Lambda_\tau, \nu_s) \alpha^2 + B_\tau(\Lambda_\tau, \nu_s) \alpha + C_\tau(\Lambda_\tau, \nu_s)
\end{equation}

\begin{equation}
A_\tau(\Lambda_\tau, \nu_s) = -10^{-2}(2.26 \nu_s^2 + 2.07 \nu_s + 2.3) \Lambda_\tau^{0.079 \nu_s^2 + 0.1754 \nu_s - 1.342}
\end{equation}

\begin{equation}
B_\tau(\Lambda_\tau, \nu_s) = (0.0573 \nu_s^2 + 0.0937 \nu_s + 0.202) \Lambda_\tau^{0.0274 \nu_s^2 + 0.0529 \nu_s - 0.8765}
\end{equation}

\begin{equation}
C_\tau(\Lambda_\tau, \nu_s) = 10^{-4}(9.654 \nu_s^2 + 4.945 \nu_s + 3.1) \Lambda_\tau^{0.01867 \nu_s^2 + 0.4011 \nu_s - 1.8186}
\end{equation}

\begin{equation}
\Lambda_n = \frac{2 G_c (1 - \nu_s)(1 - \nu_c)}{\pi G_s (1 - 2 \nu_c)}
\end{equation}

\begin{equation}
\Lambda_\tau = \frac{G_c}{\pi G_s}
\end{equation}

\begin{equation}
\alpha = \left[ \frac{\frac{2}{3} (\phi_c - \phi)}{1 - \phi_c} \right]^{0.5} 
\end{equation}

\begin{equation}
\nu_c = 0.5 \frac
{(\frac{K_c}{G_c} - \frac{2}{3})}
{(\frac{K_c}{G_c} + \frac{1}{3})}
\end{equation}

\begin{equation}
\nu_s = 0.5 \frac
{(\frac{K_s}{G_s} - \frac{2}{3} )} 
{(\frac{K_s}{G_s} + \frac{1}{3})}
\end{equation}

Phase 2: $\phi_b \rightarrow 0$
$K_b$ and $G_b$ are bulk modulus and shear modulus calculated after pre-listed equations. Bulk and shear moduli are then interpolated using HS lower bound.

\begin{equation}
K = \left[ 
\frac{\frac{\phi}{\phi_b}}
{K_b + \frac{4}{3} G_b } 
+ 
\frac{1-\frac{\phi}{\phi_b}}
{K_s + \frac{4}{3} G_b} \right]^{-1} - \frac{4}{3} G_b
\end{equation}

\begin{equation}
G = \left[ 
\frac{\frac{\phi}{\phi_b}}
{G_b + z } 
+ 
\frac{1-\frac{\phi}{\phi_b}}
{G_s + z} \right]^{-1} - z,
\end{equation}
with 
\begin{equation}
z = \frac{G_b}{6} \left( \frac{9K_b + 8\mu_b}{K_b + 2\mu_b} \right)
\end{equation}

\section{\added{Comparison with column experiment}}
\added{In this section, we describe our approach to completing the simulation and comparing the results with experimental data.\\
The sand column is simulated for the state after mineralization process, with an assumption of axial symmetric porosity throughout. Since the Unconfined Compressive Strength (UCS) test is conducted under dry conditions, fluid effects are neglected, simplifying the simulation to purely elastic behavior.\\
Given the axial symmetry of the column, we reduce the simulation to a two-dimensional domain. This 2D domain has dimensions of 25 mm radius by 100 mm height along the axis.\\
To calculate the momentum balance, we employ the following balance equation expressed in cylindrical coordinates:
\begin{equation}
\begin{aligned} 
& \frac{1}{r}\frac{\partial r \sigma_{r}}{\partial r}
 + \frac{\partial \tau_{zr}}{\partial z} 
- \frac{1}{r}\sigma_{\theta} = 0 
\\ & \frac{1}{r} \frac{\partial r \tau_{zr}}{\partial r}
 + \frac{\partial \sigma_{z}}{\partial z} = 0 ,
\end{aligned}
\label{eq: momentum symmetric}
\end{equation}
where $\sigma_{r}, \sigma_{z}$ are the normal stress along radius and axis and  $\tau_{zr}$ for the shear stress.\\
The stress-strain relationship can be determined using Hooke's Law, where the elastic moduli $E, \nu$ are determined by cementation model
\begin{equation}
\begin{pmatrix}
\sigma_r \\ \sigma_z \\ \sigma_\theta \\ \tau_{rz}
\end{pmatrix}
=
\frac{E}{(1+\nu)(1-2\nu)}
\begin{pmatrix}
1-\nu & \nu   & \nu   & 0 \\
\nu   & 1-\nu & \nu   & 0 \\
\nu   & \nu   & 1-\nu & 0 \\
0     & 0     & 0     & \frac{1-2\nu}{2}
\end{pmatrix}
\begin{pmatrix}
\varepsilon_r \\ \varepsilon_z \\ \varepsilon_\theta \\ \gamma_{rz}
\end{pmatrix},
\label{eq:axisym_stress_strain}
\end{equation}
\\
with strain given by 
\begin{equation}
\begin{pmatrix}
\varepsilon_r \\[1mm]
\varepsilon_\theta \\[1mm]
\varepsilon_z \\[1mm]
\gamma_{rz}
\end{pmatrix}
=
\begin{pmatrix}
\frac{\partial}{\partial r} & 0 \\[1mm]
\frac{1}{r} & 0 \\[1mm]
0 & \frac{\partial}{\partial z} \\[1mm]
\frac{\partial}{\partial z} & \frac{\partial}{\partial r}
\end{pmatrix}
\begin{pmatrix}
u_r \\[1mm]
u_z
\end{pmatrix}.
\label{eq:strain-displacement_axisym}
\end{equation}
The boundary condition is set as following, with a given loading velocity.
\begin{table}[ht]
\centering
\caption{Boundary Conditions for $u_r$ and $u_z$}
\label{tab:boundary_conditions}
\begin{tabular}{|c|c|c|}
\hline
\textbf{Boundary} & \textbf{\(u_r\)} & \textbf{\(u_z\)} \\ \hline
Top  & no stress & -0.045 * t\\ \hline
Bottom & no stress & 0 \\ \hline
Inner & 0 & no stress \\ \hline 
Outer & no stress & no stress \\ \hline
\end{tabular}
\end{table}\\
We calculate the average normal stress in the top cells at each time step to determine the loading pressure, allowing us to generate the stress-strain diagram shown in Fig~\ref{fig:ucs_result}.}

\section{\added{Additional results of showcase}}
\begin{figure}
    \centering
    \includegraphics[width=0.8\linewidth]{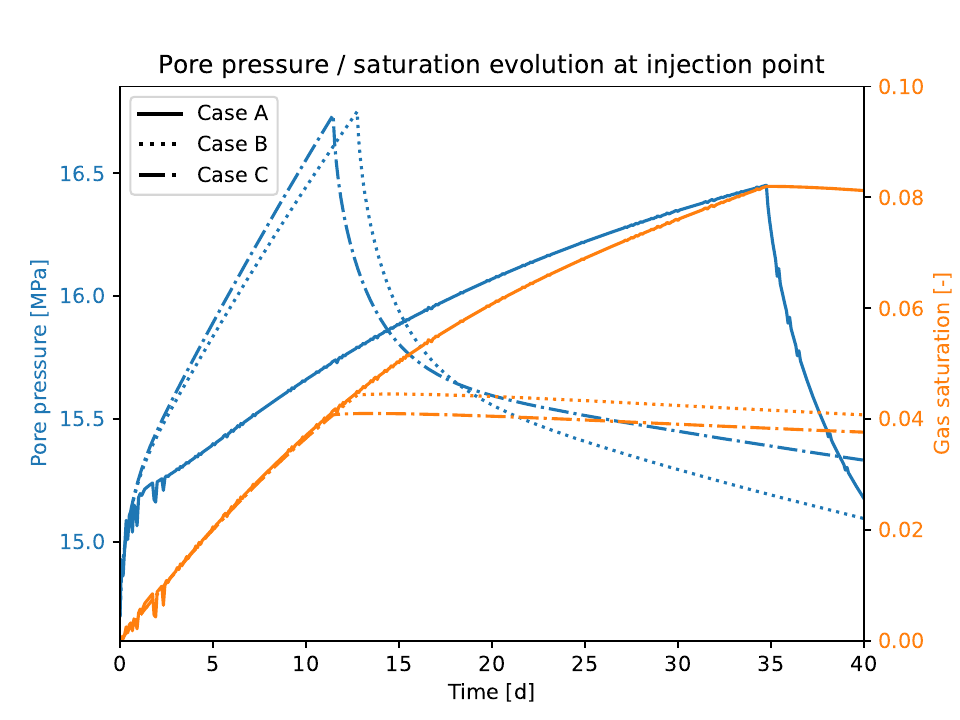}
    \caption{Time evolution of pore pressure and gas saturation in the injection zone for the three cases.}
    \label{fig:pore_pressure_evolution}
\end{figure}
\added{
Fig.~\ref{fig:pore_pressure_evolution} illustrates the evolution of pore pressure in the injection area, clearly divided into two distinct phases separated by the cessation of injection. The first phase corresponds to the injection period, characterized by a continuous increase in pore pressure. In the second phase, pore pressure gradually decreases as a result of pressure diffusion driven by fluid migration. Concurrently, gas saturation declines very slowly due to capillary effects.}

\begin{figure}
    \centering
    \includegraphics[width=0.8\linewidth]{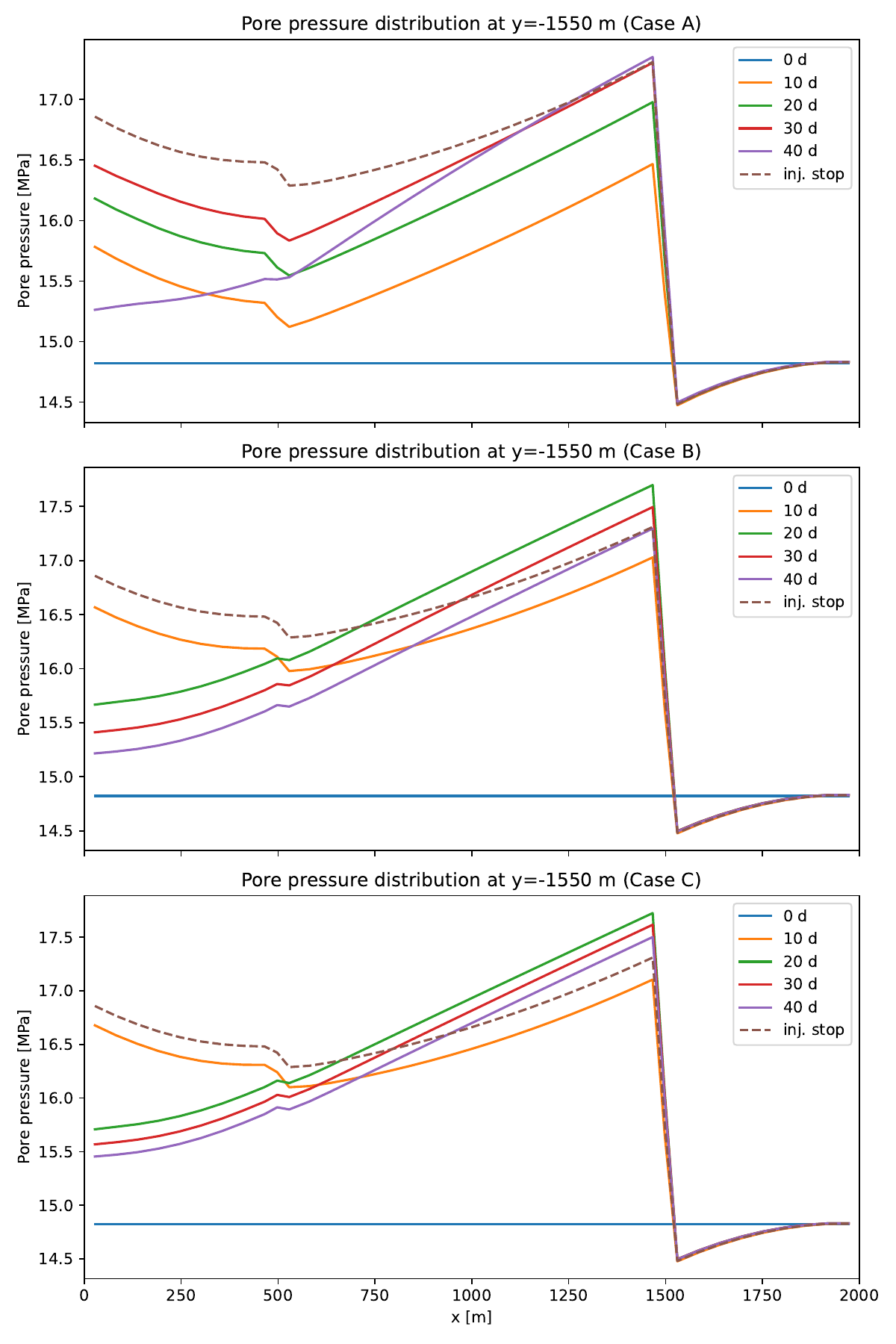}
    \caption{Pore pressure distribution at depth 1550~m  for the three cases.}
    \label{fig:pore_pressure_distribution_cases}
\end{figure}
\begin{figure}
    \centering
    \includegraphics[width=0.8\linewidth]{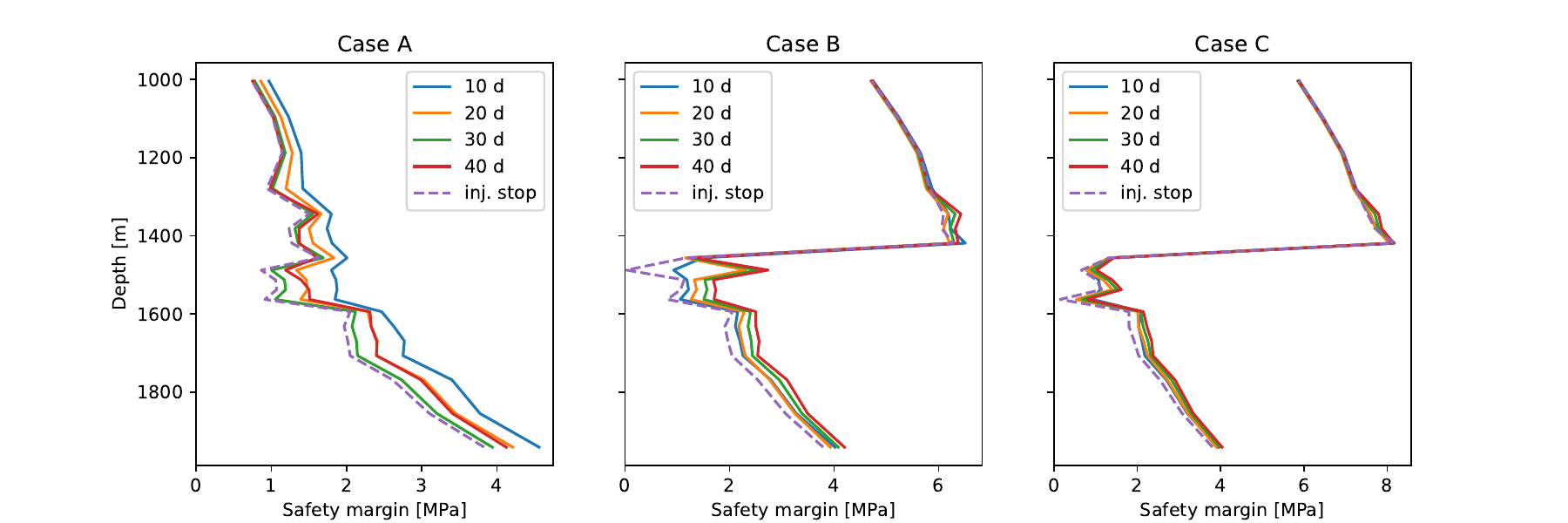}
    \caption{Time evolution of safety margin in the left fault zone for the three cases.}
    \label{fig:safety_margin_left}
\end{figure}
\begin{figure}
    \centering
    \includegraphics[width=0.8\linewidth]{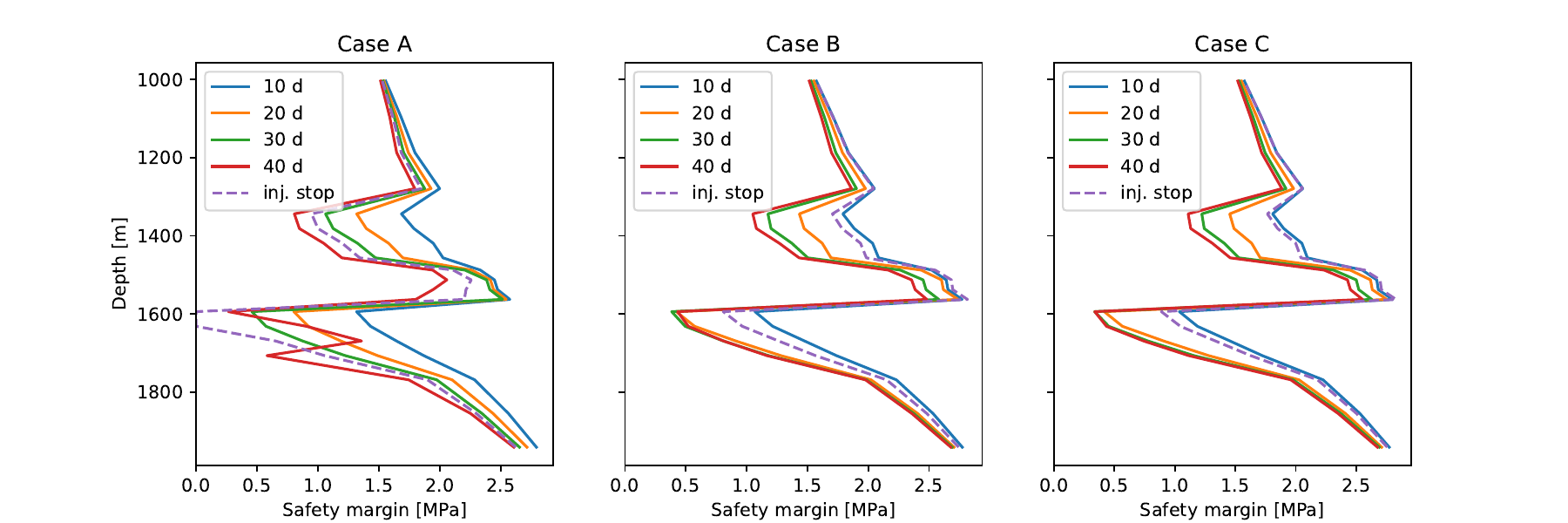}
    \caption{Time evolution of safety margin in the right fault zone for the three cases.}
    \label{fig:safety_margin_right}
\end{figure}
\begin{figure}
    \centering
    \includegraphics[width=0.8\linewidth]{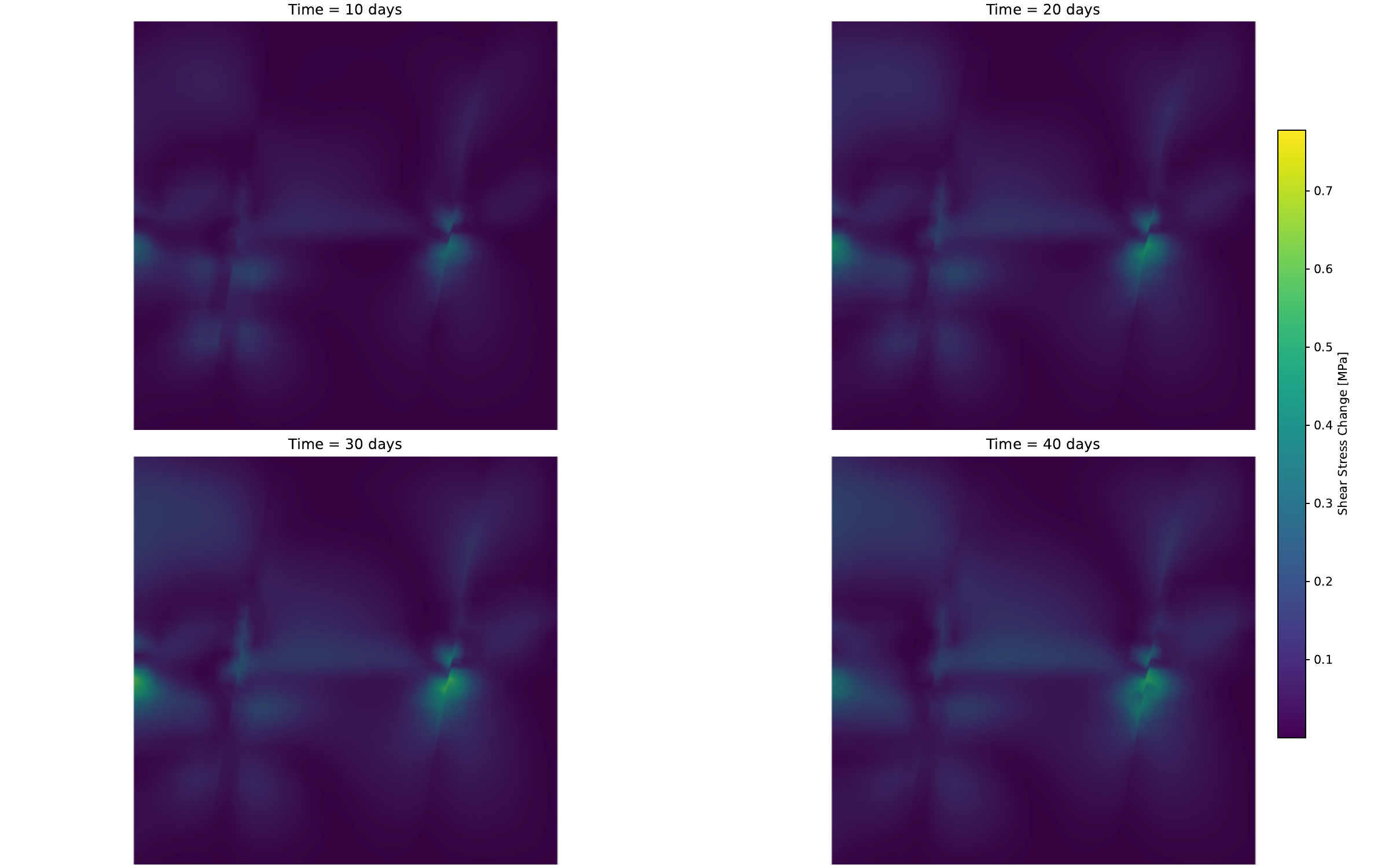}
    \caption{Time evolution of shear stress change in the domain for the three cases.}
    \label{fig:shear_stress_evolution}
\end{figure}

\added{
Fig.\ref{fig:pore_pressure_distribution_cases} illustrates the pore pressure distribution along the reservoir at a depth of 1550m for three studied cases. Two clear discontinuities in the pressure profiles highlight the influence of fault zones. The left fault, characterized by lower permeability (sealed fault), restricts fluid movement, resulting in a moderate pressure drop across this fault. Furthermore, the permeability of the fault significantly affects the rate at which pore pressure rises during injection. In contrast, beyond the second fault (on the right), the effect of injection diminishes considerably, owing to its higher permeability and proximity to a pressure boundary that enables rapid fluid drainage. The central region of each plot demonstrates a delayed pore-pressure response following injection cessation, where the pore pressure continues to rise briefly after injection ends before gradually decreasing due to pressure diffusion processes. Additionally, the slope direction observed in this central area reflects the coupled hydro-mechanical behavior: the stresses induced by injection are partially equilibrated by changes in pore pressure governed by reservoir boundary conditions.\\
The evolution of the safety margin along both fault zones is presented in Fig.\ref{fig:safety_margin_left} and Fig.\ref{fig:safety_margin_right}. The sealing (closure) condition of the left fault zone produces a distinctly different evolution pattern compared to the right fault. Both figures clearly demonstrate areas approaching shear failure, as indicated by the safety margin decreasing towards zero. Notably, regions characterized by stiffer rock properties are more prone to reaching failure conditions, underscoring the critical role of mechanical heterogeneity in controlling fault stability. Additionally, Fig.~\ref{fig:shear_stress_evolution} illustrates the variation in shear stress induced by injection for Case B, highlighting that increased shear stress predominantly develops along fault and rock zones, coinciding precisely with areas exhibiting significant mechanical heterogeneity.}

\end{appendices}
\end{document}